\RequirePackage{fix-cm}
\documentclass{svjour3}                     
\smartqed  
\usepackage{graphicx}
\usepackage{amsmath,amssymb}
\usepackage{graphicx,epsfig}
\usepackage{sidecap}
\usepackage{ragged2e}
\usepackage[usenames,dvipsnames]{color}
\usepackage[colorlinks=true,urlcolor=blue]{hyperref}
\usepackage{upgreek}
\usepackage[normalem]{ulem}
\definecolor{dgreen}{rgb}{0,0.6,0.3}
\newcommand{\CF}{{\cal F}}

\newcommand{\leftout}[1]{}
\newcommand{\sayy}[1]{`#1'}

\newcounter{thgroupcount}
\newenvironment{thgroup}{%
\setcounter{thgroupcount}{\thetheorem}
\setcounter{theorem}{0}
\refstepcounter{thgroupcount}
\renewcommand{\thetheorem}{\thethgroupcount.\alph{theorem}}
}
{\setcounter{theorem}{\thethgroupcount}}

\newcounter{propgroupcount}

\newcounter{corgroupcount}

\def\be{\begin{equation}}
\def\ee{\end{equation}}
\def\bea{\begin{eqnarray}}
\def\eea{\end{eqnarray}}
\def\lb{\label}
\def\ct{\cite}

\def\gam{\gam}

\def\eps{\epsilon}

\def\th{\theta}
\def\hth{\hat{\theta}}
\def\tith{\tilde{\theta}}
\def\sig{\sigma}
\def\hsig{\hat{\sigma}}
\def\tisig{\tilde{\sigma}}
\def\Om{\Omega}
\def\om{\omega}

\def\D{\mbox{D}}
\def\ptl{\partial}
\def\la{\langle}
\def\ra{\rangle}
\def\La{\left\langle}
\def\Ra{\right\rangle}

\def\apj{{Astrophys. J.} }

\def\cqg{{Class. Quantum Gravity} }
\def\grg{{Gen. Relativ. Gravit.} }
\def\jcap{{J. Cosmol. Astropart. Phys.} }
\def\jmp{{J. Math. Phys.} }
\def\mn{{Mon. Not. R. Astron. Soc.} }
\def\prd{{Phys. Rev.} D }
\def\prl{{Phys. Rev. Lett.} }
\def\prs{{Proc. R. Soc. Lond. A} }
\definecolor{myblue}{rgb}{0.2,0.3,0.7}
\definecolor{darkgreen}{rgb}{0,0.3,0}
\definecolor{mygreen}{rgb}{0,0.5,0}
\definecolor{grey}{rgb}{0.5,0.5,0.5}
\definecolor{orange}{rgb}{1,0.5,0}
\definecolor{MyGreen}{rgb}{0.0,.5,0.0}
\definecolor{MyDarkRed}{rgb}{0.7,0,0}
\definecolor{Pink}{rgb}{0.7,0,0.5}
\hypersetup{
    colorlinks=true,
    citecolor=blue,
    linkcolor=magenta,
    filecolor=magenta,      
    urlcolor=magenta,
}

\journalname{GRG}
\begin{document}
\title{The averaging problem on the past null cone\\ 
in inhomogeneous dust cosmologies
\thanks{Work supported by ERC advanced Grant 740021--ARTHUS.} 
}

\titlerunning{The averaging problem on the past null cone} 

\author{\mbox{Thomas~Buchert\and Henk~van~Elst\and Asta~Heinesen}}
\institute{Thomas Buchert$^{1,4,\ddagger}$ \and Henk van Elst$^{2,3}$ \and
Asta Heinesen$^{1}$ 
\at
$^1$ Univ Lyon, Ens de Lyon, Univ Lyon1, CNRS, Centre de Recherche Astrophysique de Lyon UMR5574, F--69007 Lyon, France \\ 
$^2$ parcIT GmbH, Erftstra\ss e 15, D--50672 K\"oln, Germany \\
$^{3}$ Cosmology and Gravity Group (CAGG), Department of
Mathematics and Applied Mathematics, University of Cape Town,
ZA--7701 Rondebosch, South Africa \\
$^4$ Fakult\"at f\"ur Physik, Universit\"at Bielefeld,
Postfach 100131, D--33501 Bielefeld, Germany \\
$\ddagger$ corresponding author. \\ $\ $ \\
\email{buchert@ens-lyon.fr $\cdot$ Henk.van.Elst@parcIT.de $\cdot$ asta.heinesen@ens-lyon.fr} 
}

\authorrunning{Thomas Buchert, Henk van Elst, Asta Heinesen}
\date{Received: 23 February 2022 / Accepted: 8 December 2022}

\maketitle
\begin{abstract}
Cosmological models typically neglect the complicated nature
of the spacetime manifold at small scales in order to
hypothesize idealized general relativistic solutions for
describing the average dynamics of the Universe. 
Although these solutions are remarkably successful in accounting
for data, they introduce a number of puzzles in cosmology, and
their foundational assumptions are therefore important to test.
In this paper, we go beyond the usual assumptions in cosmology
and propose a formalism for averaging the local general relativistic spacetime 
on an observer's past null cone: we formulate average properties of light fronts 
as they propagate from a cosmological emitter to an observer. 
The energy-momentum tensor is composed of an irrotational
dust source and a cosmological constant -- the
same components as in the $\Lambda$CDM model for late cosmic
times -- but the metric solution is not \emph{a priori}
constrained to be locally homogeneous or isotropic. This generally
makes the large-scale dynamics depart from that of a
simple Friedmann--Lema\^\i tre--Robertson--Walker solution
through \sayy{backreaction} effects. Our formalism quantifies such
departures through a fully covariant system of area-averaged
equations on the light fronts propagating towards an observer,
which can be directly applied to analytical and numerical
investigations of cosmic observables. For this purpose, we formulate light front averages of
observable quantities, including the effective
angular diameter distance and the cosmological redshift drift and we also discuss 
the backreaction effects for these observables. 

\keywords{Relativistic cosmology \and Spacetime foliations \and
Light propagation \and Cosmological backreaction \and Dark
Universe}
\end{abstract}
\clearpage
\setcounter{tocdepth}{3}
\normalsize\tableofcontents

\section{Introduction}
Observational cosmology primarily relies on information that
propagates along our past null cone; \textit{cf.}
Ref.~\cite{krisac1966}. 
Observational data is most commonly interpreted within the
Friedmann--Lema\^\i tre--Robertson--Walker
(FLRW) class of spacetime metrics. Perturbation theory at this
background cosmology is then employed to refine the description of
the propagation of null signals along the past null cone in an
`almost FLRW' cosmology that takes inhomogeneities into account
(see, e.g., Refs.
\cite{bildhauer:backreaction2,sasaki1,sasaki3,vanderveldetal}). 
The $\Lambda$CDM model (Cold Dark Matter including a positive 
cosmological constant $\Lambda$) is a particular general
relativistic FLRW metric model, with ordinary matter, radiation,
dark matter (CDM), and dark energy (modelled with
$\Lambda$) as energy-momentum sources. The $\Lambda$CDM paradigm
has proven successful in a largely consistent interpretation of
observational data; however, with a number of `tensions' remaining
\cite{FLRWtensions,kids,Riess2019,hubbletension1,hubbletension2,Perivolaropoulos:2021jda}. 
Importantly, the success
of the $\Lambda$CDM framework comes at the price of introducing
dark components to the cosmological energy budget \cite{planck}.

The specification of a realistic inhomogeneous spacetime metric for
a lumpy model of the Universe and the description of
realistic local properties for the propagation of null signals 
is an involved task. Cosmological observations involve the study of
many sources and the propagation of null signals over large
spatial distances. We might thus aim at formulating average
dynamical equations -- suitable for obtaining effective laws for
the propagation of null signals in a cosmological context -- where
exact knowledge of the local spacetime metric might not be
required. Averaging of Einstein's field equations in general
relativity has been formulated in the context of a $3$-dimensional
volume-averaging of scalar-valued variables\footnote{Averaging over
tensor-valued variables introduces ambiguities in the
volume-averaging procedure (see
Refs.~\cite{ellis,ellisbuchert,buchert:review,klingon,singh2}
and references therein).}
\cite{buchert:grgdust,buchert:grgfluid}. In this approach,
the spacetime metric is not specified and
no approximation is employed to restrict inhomogeneities.
Instead, one concentrates on integral properties of local
spacetime variables such as  the rest mass density, the
expansion rate, the shear rate and the spatial Ricci
curvature scalar. The resulting volume-averaged Einstein's field
equations can be viewed as balance equations for effective macroscopic  
spacetime variables. The system of equations governing the macroscopic 
variables is in general not closed and 
additional constraints must be imposed in order to solve the
system. (A comprehensive discussion
of the closure problem in $N$ dimensions, in particular
emphasizing topological constraints, may be found in
Ref.~\cite{GBC}.)

In such average Universe descriptions, the structure of the
underlying local inhomogeneous distributions of the matter
content and the spatial geometry manifests itself globally through 
correction terms to Friedmann's equations. Such terms will be
non-vanishing in general -- also in cosmologies with a notion of
statistical spatial homogeneity and isotropy -- and
represent the deviance of the \sayy{monopole state} of a
model universe with structure, i.e.,
an appropriate lowest-order and spatially uniform solution,  from the strictly spatially
homogeneous and isotropic FLRW solutions. 
The monopole state can,
nevertheless, be mapped to an effective FLRW cosmology by
interpreting the correction terms as effective source terms. For
instance, volume-averaged fluctuation terms have been
discussed as geometrical candidates of kinematical dark energy and
dark matter sources; e.g. Refs.
\cite{buchert:jgrg,palle,morphon,David1,buchert:darkenergy,kolbetal,rasanen,lischwarz,buchert:review,Bolejko1,David2,wiegandbuchert,RZA_2,roy:instability,sussman1,Buchert2011,Rasanenreview,buchertrasanen,RZA_2,sussman2,David3,zimdahl:LTB,Bolejko2,quentin:darkmatter,hubbletension3}. 

Much work has been done in describing the past null cones of observers in various
inhomogeneous and anisotropic model universes (e.g. Refs.
\cite{newpen1962,dyeroe74,dyeroe75,weinberg,linsenbuch,maamat1994,mustapha1,ellis:dahlem,kantowski3,kantowski2,kantowski1,demetal03,pyne,Gourgoulhon2006,ruth:luminosity,hellaby:mass,luhellaby,dominik:average}),
including coordinate-indepen\-dent considerations within gauge invariant perturbation theory \cite{Marco:perturbations,YooDurrer,Yoo:2019qsl}, and the distortion of the average distance--redshift relation
relative to that of a reference FLRW cosmology in specific
spatially inhomogeneous model universes \cite{mustapha2,SikoraGlod,LTBluminosity1,LTBluminosity2,LTBluminosity3,Fleury:2013sna,Fleury:2013uqa,Koksbang:2017arw,Adamek:2018rru}. 
General considerations on the propagation of null signals
in statistically spatially homogeneous and isotropic cosmologies
with slowly evolving structure have been carried out
\cite{lightpropndust,lightpropngeneral}.
Such considerations allow for relating leading-order
observable quantities to volume-averaged variables on spacelike
$3$-surfaces under the assumption that null signals 
sample the Universe fairly in volume. 
Covariant null cone averaging formalisms suitable for
averaging on light fronts have been developed in
Refs.~\cite{lightconeav1,Fanizza2019pfp}, but
have mainly been employed for the purpose of constraining
errors in the determination of FLRW background parameters in a
$\Lambda$CDM cosmology \cite{lightconeav2,BenDayan:2012,BenDayan:2013gc}.
Formalism that can be used for model-independent analysis of
standardizable objects and cosmic drift effects, while making no
assumptions about the form of the cosmological spacetime
metric, has been presented in Refs.~\cite{Umeh:2013UCT,Heinesen:2020bej,Heinesen:2021nrc,Heinesen:2021qnl} -- see also Refs.~\cite{Nez,Korzynski:2017nas,Grasso:2018mei,Korzynski:2021aqk,carfora} for
other interesting formalisms which could be suitable for model-independent data analysis.

In this paper, we employ an averaging operation similar 
to that introduced
in Refs.~\cite{buchert:grgdust,buchert:grgfluid},\footnote{The
same averaging operation might be formulated in the notation of
Refs.~\cite{lightconeav1,Fanizza2019pfp}.} but with the averaging
domain adapted to an observer's past null cone.
More precisely, we shall propagate $2$-dimensional inhomogeneous
light fronts on which we define area-averaged
physical variables (of, e.g., matter density and optical scalars), 
and invoke Einstein's field equations to obtain propagation
equations for the same area-averaged 
variables. We do this \textit{without} specifying the spacetime
metric. We shall characterize average cosmological distance
relations in terms of null cone averages over light fronts.
We shall constrain the local setting to that of an irrotational
dust cosmology. 
This approximation comes with limitations: on scales where
gravitationally bound structures are forming, the
irrotationality assumption for matter breaks down. At such
scales, velocity dispersion arising from the internal motion
effects within the structures must generally also be included
through an effective anisotropic pressure term that violates the dust
approximation. In the early history of the Universe, where
radiation constitutes a significant fraction of the total energy
density, pressure must also be included for an accurate
description. The framework developed here is thus applicable for
scales above the largest gravitationally virialized structures and for
cosmic times well after the epoch of radiation and matter
equality. In order to include modelling of smaller length scales and
earlier cosmic times, the framework must be generalized in terms
of the assumed matter content, which may be done by following
similar approaches as in Ref.~\cite{buchertetal:generalfluid}, see also
\cite{buchertetal:foliations}.
We furthermore consider the
idealized case where caustics in the geodesic null congruence
constituting an observer's past null cone can be ignored. 
This is consistent with the large-scale modelling of the
matter content of the Universe described above, where we neglect
the physics of strong gravitational lenses. 
For treatments including caustics, see, e.g.,
Refs.~\cite{ellissolomons,mustapha3}. 

The general framework
investigated in this paper offers the possibility of employing
non-perturbative and background-free approximations to describe
the effect of inhomogeneities on cosmological measurements. 

\bigskip
\noindent
\underbar{Notation and conventions:} 
We use units in which
$c=G/c^{2}=1$. Greek letters $\mu, \nu, \ldots$ label
spacetime indices in a general basis. Summation over repeated
indices is understood. The signature of the spacetime metric,
$g_{\mu \nu}$, is taken to be $(- + + +)$ \cite{dreimaennerbuch}.

\section{Local spacetime setting}
Here we describe the assumptions made for the local
cosmological spacetime. In section \ref{sec:fieldeq} we describe 
the assumed matter content in the form of an irrotational dust
source. In section \ref{sec:nullcone} we describe the observer's
past null cone and in section \ref{sec:foliations} we introduce
the foliation of the past null cone into light fronts that will
be used throughout this paper. In sections \ref{sec:prop} and
\ref{sec:constraints} we derive propagation equations and
constraint equations, respectively, for capturing the dynamics
of relevant variables for the cosmological spacetime and null
signals travelling along an observer's past null cone.
Finally, in section \ref{sec:conservation} we consider photon
number conservation and effective cosmological distance measures. 

\subsection{Field equations and matter content}
\label{sec:fieldeq}
We consider a cosmological spacetime which is dynamically
described through Einstein's field equations \ct{ein1915,ein1917},
\be
\lb{einstein}
R_{\mu \nu} - \frac{1}{2}\,g_{\mu \nu}R + \Lambda g_{\mu \nu}
= 8\pi T_{\mu \nu} \ ,
\ee
where $R_{\mu \nu}$ is the $4$-Ricci curvature of the spacetime, 
$R$ is its trace, and $\Lambda$ denotes the 
cosmological constant. We concentrate on a
matter-dominated epoch of the cosmological evolution and
restrict the energy-momentum tensor $T_{\mu \nu}$ to that of 
an irrotational continuum of dust,
\bea \label{eq:dust}
& & T_{\mu \nu } = \varrho u_{\mu }u_{\nu } \ ;
\quad  \om_{\mu \nu }: = h_{[\mu}{}^{\alpha}
h_{\nu]}{}^{\beta} \nabla_{ \alpha }u_{\beta } = 0 \ ,
\eea 
where $u^{\mu}$ is the future-directed $4$-velocity of the
matter congruence, $\varrho$ is the matter's rest mass density,
$\nabla_\mu $ is the covariant derivative, and square
brackets denote anti-symmetri\-zation in the indices involved.
We assume that the irrotational continuum of dust is
transparent to propagating null signals. 
From Eqs.~(\ref{eq:dust}) and the conservation of energy-momentum
it follows that $u^\mu$ is geodesic and can be expressed as the
gradient of a scalar-valued variable,
\bea
& u^{\nu}\nabla_{\nu}u^\mu = 0 \ ; \quad
u_{\mu} = - \,\partial_{\mu} \tau \ ,
\eea
where $\tau$ can be interpreted as a proper time function along
each individual matter world line, since
$u^{\rho}\partial_{\rho}\tau = 1$. 
The covariant $(1+3)$-decomposition of the covariant
derivative of the $4$-velocity of the matter congruence thus
results in (\textit{cf.} Refs.~\ct{ehl1961,ell1971,hve1996,wainwrightellis,hveetal1997,hve1998,ellhve1999,elletal2012,roy:1+3})
\be
\lb{udec}
\nabla_\mu u_\nu  = \Theta_{\mu \nu}
= \frac{1}{3}\,\th h_{\mu \nu} + \sig_{\mu \nu} \ ,
\ee
where $\th$ is the isotropic expansion rate
describing the volume expansion of the matter congruence, and
$\sig_{\mu \nu}$ is the volume shear rate describing the
volume-preserving deformation of the congruence. The covariant
derivative of the $4$-velocity is equal to its own projection
$\Theta_{\mu \nu} :=  h_\mu {}^\alpha  h_\nu {}^\beta 
\nabla_\alpha u_\beta  $ because of the vanishing of the
$4$-acceleration of $u^\mu$. The spatial projection tensor 
\be
\lb{spaceprojection}
h_\mu {}^\nu := \delta_\mu {}^\nu + u_\mu u^\nu \ ,
\ee
is orthogonal to the irrotational matter $4$-velocity field
and plays the role of the metric tensor on the spacelike
$3$-surfaces orthogonal to $u^{\mu}$. 

\subsection{Past null cone} 
\label{sec:nullcone}
We now assume the presence of a non-gravitating, geodesic and
irrotational null congruence pervading the cosmological 
spacetime such that its future-directed $4$-momentum
$k^\mu$ obeys \cite{Seitz:1994xf}
\bea
\label{eq:kproperties}
&k_\mu k^\mu = 0  \ ; \quad  k^\nu \nabla_\nu k^\mu = 0 \ ; \quad   
k_{[\mu} \nabla_{  \nu }k_{\alpha ] } = 0 \ .
\eea
This implies that $k^\mu$ is proportional to the gradient of a
scalar-valued variable, 
\bea \label{eq:kproperties2}
& k_\mu = \eta \,\partial_\mu V \ ;  \qquad 
k^{\rho}\partial_{\rho}V = 0 \ ;  \quad 
k^{\rho}\partial_{\rho}\eta  = 0 \ , 
\eea
where the conditions $\eta > 0$ and $n^{\rho}\partial_{\rho}V < 0$
(with $n^\mu$ a future-directed timelike vector field) 
ensure the future-directed nature of $k^\mu$. The function $V$
defines the direction of the $4$-momentum $k^\mu$ and $\eta$
defines its normalization. The preservation of $V$ along each
individual ray of the geodesic null congruence follows from
the vanishing norm of $k^\mu$, and the preservation of $\eta$
follows from the affine geodesic equation. 
Spacetime domains singled out by
$\{V = \text{constant}\}$ are $3$-dimensional null surfaces. 

As a special case of null surfaces we consider those
associated with the past null cones of an observer. 
We consider the case where the timelike world line,
$\gamma$, of this observer belongs to the matter
congruence.\footnote{We might in principle choose any world line
to initialize a past null cone, but we shall often be interested
in observers comoving with the matter in the cosmological
spacetime.} For each point $P$ of $\gamma$, we
consider incoming geodesic null rays of all spatial
directions orthogonal to $u^\mu$ as initial conditions for the
direction of the $4$-momentum $k^\mu$. We extend this two-parameter
family of null directions away from $P$ via the geodesic equation
to construct the past null cone ${\cal C}^{-}(P)$ at $P$. 
From $u^{\rho}\partial_{\rho}V < 0$, we have that the scalar $V$
uniquely labels the one-parameter family of past null cones
${\cal C}^{-}(P)$ along $\gamma$. Since $\eta$ is a constant
over each null cone, we have that $\eta$ is a function of $V$, and 
can be absorbed into a redefinition of $V$. We thus set
$\eta = 1$ without loss of generality. The normalization of the
$4$-momentum $k^\mu $ might for instance be chosen by fixing the
energy function,
\be
E := - u^\mu k_\mu = - u^{\rho}\partial_{\rho}V \ , 
\ee
along $\gamma$. This choice of normalization specifies the
frequency of monochromatic null signals 
measured at the observer's telescope. 
We might for instance consider the normalization
$- u^\mu k_\mu \lvert_\gamma = \text{constant}$ that corresponds
to the situation where the central observer performs measurements
at the same radiation frequency at different instances of proper
time.

The cosmological redshift, $z$, of a luminous astrophysical source
comoving with the matter congruence and located on the observer's
past null cone, as measured at the point of observation $P$
(``here and now''),  
 is given by (\textit{cf.} Ref.~\ct{ellhve1999})
\be
\lb{redshift}
(1+z) := \frac{-u^\mu k_{\mu}}{-u^\mu k_{\mu}(P)}
= \frac{E}{E(P)} \ ,
\ee
where $E(P)$ denotes the energy of the incoming null ray as
measured by the observer at the point of observation $P$. The
redshift function is
independent of the normalization procedure chosen for $k^\mu$. 

We may decompose the $4$-momentum~$k^\mu$ associated with the
incoming geodesic null congruence in terms of the $4$-velocity
$u^\mu$ of the matter congruence and a spatial unit vector $e^\mu$ 
orthogonal to $u^\mu$ according to
\begin{subequations}
\be
\lb{def:kdef}
k^{\mu} = E(u^{\mu}-e^{\mu}) \ ,
\ee 
where $e^\mu$, as evaluated by the observer at $P$, is
multi-valued and takes values in the two-parameter family of all
possible spatial directions of pointings of the observer's
telescope.  For later reference, we introduce an auxiliary
future-directed null $4$-momentum,
\be
\lb{def:ldef}
l^{\mu} = E(u^{\mu}+e^{\mu}) \ .
\ee
\end{subequations}
The null $4$-momentum $l^{\mu}$ points in the opposite spatial
direction as $k^{\mu}$, when viewed in the rest frame of the
matter congruence. Note that $l^{\mu}$ is not necessarily a
geodesic null congruence (\textit{cf.} footnote~\ref{footlmu} for
further properties).  
We refer to geodesic null rays generated
by $k^{\mu}$ as \textit{incoming null rays} and 
null rays generated by $l^{\mu}$ as \textit{outgoing null rays}. 
The field $l^{\mu}$ does not incorporate any new physical
information about the cosmological spacetime, but will be
convenient for formulating constraint equations in
the covariant $(1+1+2)$-decomposition of the cosmological
spacetime considered later in this analysis
(see Sec.~\ref{sec:constraints}).
 
\subsection{Foliation of the past null cone and light fronts}
\lb{sec:foliations}
In order to describe the evolution along a fixed past null cone
${\cal C}^{-}(P):\!\{V=\text{constant}\}$, we consider its 
foliation into a union of spatial
$2$-surfaces that we refer to as \textit{light fronts}. One way to
construct light fronts is to consider the intersection of the past
null cone with a family of $3$-surfaces (timelike, spacelike
or lightlike). 
An obvious choice of an intersecting foliation is that defined
from the irrotational $4$-velocity $u^\mu$ of the matter
congruence, so that the fronts of intersection are simultaneously
orthogonal to the null congruence and to the matter congruence. 

The resulting light fronts -- also referred to as
\textit{screen spaces}, \textit{cf.} Refs. \cite{sac1961,wal1984}
-- are constant-level surfaces of $\tau$ and $V$. 
The local screen basis propagated along the geodesics null rays is
usually referred to as \textit{Sachs basis}
\cite{sac1961,per2004,joretal1961}. In this setting,
the full observer's past null cone can be considered to be the
union of $2$-dimensional light fronts with varying values of
$\tau$.\footnote{Another option is to consider level surfaces of
constant affine parameter $\lambda$ of the geodesic null
congruence defined from the propagation requirement
$k^{\rho}\partial_{\rho}\lambda = 1$. In order to uniquely define
$\lambda$ as a spacetime function, we must specify initial
conditions. We might, for instance, require setting
$\lambda \lvert_\gamma = 0$. From this it follows immediately
that $u^{\rho}\partial_{\rho}\lambda \lvert_\gamma = 0$, and the
gradient of $\lambda$ is thus spacelike in the vicinity of
$\gamma$, and $\lambda = \text{constant}$-level surfaces define
timelike cylinders in the same vicinity. Far away from the vertex,
the use of $\lambda$ as a meaningful foliation scalar must be
carefully re-assessed.}
We define the projection tensor onto the light fronts:
\be
\lb{projection}
p_{\mu}{}^{\nu} := \delta_{\mu}{}^{\nu}
+ u_{\mu} u^{\nu} - e_{\mu}e^{\nu}  \ .
\ee
We have $p_{\mu}{}^{\nu}u_{\nu} = p_{\mu}{}^{\nu}e_{\nu} =
p_{\mu}{}^{\nu}k_{\nu} = p_{\mu}{}^{\nu}l_{\nu} = 0$,
$p_{\mu}{}^{\alpha}p_{\alpha}{}^{\nu} = p_{\mu}{}^{\nu}$, and
$p_{\mu}{}^{\mu} = 2$. 
The tensor $p_{\mu}{}^{\nu}$ is thus orthogonal to the space
spanned by   $k^\mu$ and $u^\mu$ while it acts as the
metric tensor for tensorial fields intrinsic to the screen space. 
The surface density on the light fronts is given by
\be
\lb{areaelem}
\eps_{\mu\nu} :=
u^{\alpha}p_{\mu}{}^{\beta}p_{\nu}{}^{\gamma}e^{\delta}
\eps_{\alpha\beta\gamma\delta}
=\eps_{[\mu\nu]} \ ,
\ee
where $\eps_{\alpha\beta\gamma\delta}$ is the spacetime 
permutation tensor, equal to $\sqrt{-g}$ for even and $-\sqrt{-g}$
for odd permutations of $0,1,2,3$, and $g$ is the determinant of
the cosmological spacetime metric. The surface density has the
following properties: 
$\eps_{\mu\nu}u^{\nu} = \eps_{\mu\nu}e^{\nu}
= \eps_{\mu\nu}k^{\nu} =
\eps_{\mu\nu}l^{\nu} = 0$, $\eps_{\mu\nu}\eps^{\alpha\beta} =
2!\,p_{[\mu}{}^{\alpha}p_{\nu]}{}^{\beta}$
and $\eps_{\mu\nu}\eps^{\mu\nu} = 2$. 
The area expansion tensor associated with the light fronts can
be decomposed as \cite{sac1961,joretal1961} 
\begin{subequations}
\bea
\lb{def:optscalin}
\hat{\Theta}_{\mu \nu}
:= p_{\mu}{}^{\alpha}p_{\nu}{}^{\beta}\nabla_{\alpha}k_{\beta}
& = & \frac{1}{2}\,\hth p_{\mu\nu}+\hsig_{\mu\nu} \ ,
\eea
where $\hth := p^{\mu \nu}\nabla_{\mu}k_{\nu}$ is the
dimensionless area expansion rate of a geodesic null ray bundle
and $\hsig_{\mu \nu} := p_{ \mu }{}^{\alpha }p_{\nu }{}^{\beta}
\nabla_{\alpha}k_{\beta} - \frac{1}{2}\hth p_{\mu \nu }$ is the
dimensionless area-preserving deformation or \sayy{shearing} rate
of the same geodesic null ray bundle. The anti-symmetric part of
the deformation vanishes by construction via the requirement of
irrotationality; see the last condition in
Eq.~(\ref{eq:kproperties}). We define an analogous area expansion
tensor for the auxiliary null $4$-momentum $l^{\mu}$ by 
\bea
\lb{def:optscalout}
\tilde{\Theta}_{\mu \nu}
:=  p_{\mu}{}^{\alpha}p_{\nu}{}^{\beta}\nabla_{\alpha}l_{\beta} 
& = & \frac{1}{2}\,\tith p_{\mu \nu}+\tisig_{\mu \nu} \ ,    
\eea
where we have labelled the kinematic variables associated with
$k^{\mu}$ by a hat and the kinematic variables associated
with $l^{\mu}$ by a tilde.\footnote{\label{footlmu} As a result of
the fact that $l^\mu$ does not in general generate a geodesic null
congruence, it does not in general satisfy the
condition (\ref{eq:kproperties}). For the same reason, $l^\mu$ is
not necessarily hypersurface-forming. However, it is irrotational
through its definition as a linear combination of irrotational
vector fields, Eq.~(\ref{def:ldef}), after projection onto the
screen space normal to $k^\mu$ and $u^\mu$:
$p_{\mu}^{\ \alpha} \; p_{\nu}^{\ \beta} \nabla_{[\alpha}
l_{\beta]} = 0$.}
 
It shall furthermore be useful to also define the analogous
expansion tensor and kinematic variables associated with $e^\mu$
(labelled by a bar):
\be
\lb{edec}
\overline{\Theta}_{\mu \nu } =  p_{\mu }{}^{\alpha}
p_{\nu }{}^{\beta }\nabla_{\alpha}e_{\beta}
= \frac{1}{2}\,\overline{\theta} \, p_{\mu \nu}
+ \overline{\sigma}_{\mu \nu} \ .
\ee
\end{subequations}
The trace component of $\overline{\Theta}_{\mu \nu }$ is given
by $\overline{\theta} :=
p^{\mu \nu }\nabla_{\mu}e_{\nu} = \nabla_{\mu}e^{\mu}$, which
follows from the orthonormality of $u^\mu$ and $e^\mu$ and the
geodesic nature of $u^\mu$. The symmetric tracefree part of
$\overline{\Theta}_{\mu \nu }$ is $\overline{\sigma}_{\mu\nu}
:=  p_{(\mu}{}^{\alpha}p_{\nu)}{}^{\beta}
\nabla_{\alpha}e_{\beta} - \frac{1}{2}\,\overline{\theta}\,
p_{\mu \nu}$. The anti-symmetric component
$p_{[\mu}{}^{\alpha}p_{\nu]}{}^{\beta} \nabla_{\alpha}e_{\beta}$
vanishes due to the requirement of irrotationality
for both $u^\mu$ and $k^\mu$. 

The evolution of the multi-valued spatial direction vector
$e^\mu$ along the geodesic null congruence generated by $k^{\mu}$
can be formulated as \cite{Heinesen:2020bej}
\bea
\label{kderive}
E^{-1} k^\rho \nabla_\rho  e^\mu
= - E^{-1} \alpha  k^\mu - \left(\frac{1}{3}\,\theta
h_{\nu}{}^{\mu} + \sigma_{\nu}{}^{\mu}\right)e^{\nu} \ , 
\eea
where we have made use of one of the two area-adapted longitudinal
expansion rate variables for the matter congruence,
\be
\lb{eq:alphbet}
\alpha := \frac{1}{3}\,\th - (p^{\mu\nu}\sig_{\mu\nu}) \ ,
\qquad
\beta := \frac{1}{3}\,\th
+\frac{1}{2}\,(p^{\mu\nu}\sig_{\mu\nu}) \ ,
\ee
which were introduced in Ref.~\ct{steell1968}. 
In the present irrotational dust cosmology setup the variable
$\alpha$ determines the logarithmic rate of change of the
energy $E$ along the geodesic null rays, see Sec.~\ref{sec:prop},
and replaces the Hubble parameter in FLRW cosmology in the
general cosmographic representation of the luminosity distance;
see Refs.~\cite{Umeh:2013UCT,Clarkson:2011uk,Heinesen:2020bej}.
Using Eq.~(\ref{kderive}), we can write the \sayy{acceleration}
vector of $e^\mu$ in the following way:
\bea
\lb{drewrite}
h_{\nu}{}^{\mu} e^\alpha \nabla_\alpha e^\nu 
= \kappa^\mu  + p_{\nu}{}^{\mu}e^{\rho}\sigma_{\rho}{}^{\nu} \ ,
\eea
where $\kappa^{\mu} := p_{\nu}{}^{\mu}u^{\rho}\nabla_{\rho}e^{\nu}$
is the position drift describing the angular drift of the source
on the observer's sky as seen in a non-rotating (Fermi-propagated)
reference frame \cite{Korzynski:2017nas}. The acceleration 
vector (\ref{drewrite}) involves the volume shear rate and the
angular drift of the source on the observer's sky and can thus be
thought of as a measure of the violation of isotropy around the
observer's position. The acceleration vector (\ref{drewrite}) and
$\kappa^{\mu}$ are both constructed such that they are
tangential to the $2$-dimensional screen space.

The dimensionless kinematic variables $\hth$,
$\hsig_{\mu\nu}$, $\tith$ and $\tisig_{\mu \nu}$ associated with
the null $4$-momenta $k^{\mu}$ and $l^{\mu}$ can be expressed
algebraically in terms of the kinematic variables of $u^\mu$ and 
$e^\mu$, and the energy $E$ of the incoming null rays as follows:
\begin{subequations}
\bea
\lb{kinhat}
\frac{\hth}{E} = 2\beta - \overline{\theta} \ , \qquad
\frac{\hsig_{\mu\nu}}{E} =
p_{\la \mu}{}^{\alpha}p_{\nu\ra}{}^{\beta}\sig_{\alpha\beta}
- \overline{\sigma}_{\mu\nu} \ , \\
\lb{kintilde}
\frac{\tith}{E} = 2\beta + \overline{\theta} \ , \qquad
\frac{\tisig_{\mu \nu}}{E} = 
p_{\la \mu}{}^{\alpha}p_{\nu\ra}{}^{\beta}\sig_{\alpha\beta}
+ \overline{\sigma}_{\mu\nu} \ ;
\eea 
\end{subequations}
this result is obtained from the covariant decompositions
(\ref{def:kdef}) and (\ref{def:ldef}), and the definitions of the
kinematic variables introduced in Eqs.~(\ref{udec}),
(\ref{def:optscalin}), (\ref{def:optscalout}) and (\ref{edec}).
Angular brackets single out the symmetric and tracefree
part of the tensor in the involved indices.

The null kinematic variables $\hth$, $\hsig_{\mu\nu}$,
$\tith$ and $\tisig_{\mu \nu}$ inherit the kinematics of $u^\mu$
and $e^\mu$ in the screen space. The longitudinal component of the
shear rate, $(p^{\mu \nu}\sig_{\mu \nu})$, enters in the
expansion rates $\hth$ and $\tith$ via $\beta$. In general, the
kinematic variables of $k^\mu$ and $l^\mu$ differ. This difference
can be assigned to the focusing of the null congruences towards
the observer, and is for instance present for radially
propagating geodesic null congruences in FLRW cosmology.

We may eliminate the kinematic variables of the matter
congruence in Eqs.~(\ref{kinhat}) and (\ref{kintilde}) to obtain
the useful relations:
\be
\lb{inout}
\tilde{\theta} = \hth + 2 E \overline{\theta} \ , \qquad
\quad \tilde{\sigma}_{\mu\nu} = \hat{\sigma}_{\mu\nu}
+ 2 E \overline{\sigma}_{\mu\nu} \ . 
\ee
The function $\overline{\theta}$ quantifies the departure of the
$2$-dimensional screen space from a minimal surface
area\footnote{It is a well-known result from calculus of variations
in Riemannian geometry that surfaces minimizing the area measure
locally have zero trace of the extrinsic curvature scalar of the
embedding.} within the spacelike $3$-surfaces orthogonal to
$u^\mu$. When $k^\mu$ is a generator of a past null cone, which is
narrowing towards the singularity at the vertex of the observer, 
we expect $\overline{\theta}$
to be dominantly negative.\footnote{Since causal lines can only
leave a past null cone (and not enter), we indeed expect a
negative contribution to the overall expansion rate of the screen
space from the drift of the screen space boundaries
relative to the matter congruence: the screen space is
sampling the cross section of fewer fluid elements as the vertex
of the past null cone is approached. However, local differential
expansion of the dust matter congruence and the spatial
fluctuations in the rest mass density $\varrho$ could
potentially compensate this tendency locally.}

\subsection{Propagation equations along the past null cone}
\lb{sec:prop}
We now formulate the evolution equations along the observer's
past null cone ${\cal C}^{-}(P):\!\{V = \text{constant}\}$
for the variables of our main interest.

The propagation equations
for the optical scalars associated with the incoming geodesic
null congruence $k^{\mu}$ can be derived from
light front projections of the Ricci identity $2\nabla_{[\mu }
\nabla_{\nu ]}k^\alpha =-R_{\mu \nu \beta}{}^\alpha k^\beta $;
\textit{cf.} Refs.~\ct[pp.~222--223]{wal1984} and \ct{joretal1961}.
Evolving the energy-rescaled variables of Eq.~(\ref{kinhat}),
we obtain:
\begin{subequations}
\bea
\lb{opticalenergy}
&&E^{-1}k^{\rho}\partial_{\rho}E
 =  -\,\alpha E \ ; \\
\lb{opticalraychaudhuri}
&&E^{-1}k^{\rho}\partial_{\rho}\left(\frac{\hth}{E}\right)
=  -\frac{1}{2}\left(\frac{\hth}{E}-2\alpha\right)
\left(\frac{\hth}{E}\right) - 2\left(\frac{\hsig}{E}\right)^{2}
- 8\pi\varrho \ ; \\
\lb{opticalshear}
&&E^{-1}k^{\rho}\partial_{\rho}\left(\frac{\hsig}{E}\right)^{2}
 =  -2\left(\frac{\hth}{E}-\alpha\right)
\left(\frac{\hsig}{E}\right)^{2} \nonumber\\
&&\qquad\qquad\qquad\qquad\
-\ 2\left(\frac{\hsig^{\mu \nu}}{E}\right)\left(
p_{\la \mu}{}^{\alpha}p_{\nu\ra}{}^{\beta}E_{\alpha\beta}
+\eps_{\la \mu}{}^{\alpha}p_{\nu\ra}{}^{\beta}H_{\alpha\beta} 
\right) \ ,
\eea
\end{subequations}
where $\hsig^{2} := \frac{1}{2}\hsig_{\mu\nu}\hsig^{\mu\nu}$
is the (squared) optical shear scalar. 
In Eq.~(\ref{opticalshear}), the propagation
of the energy-rescaled optical shear scalar is sourced by the
incoming radiative Weyl curvature eigenfield defined by
\be
\lb{weylin}
p_{\la \mu}{}^{\alpha}p_{\nu\ra}{}^{\beta}E_{\alpha\beta}
+\eps_{\la \mu}{}^{\alpha}p_{\nu\ra}{}^{\beta}H_{\alpha\beta} \ ,
\ee
where $E_{\alpha\beta} := C_{\mu\nu\rho\sigma}
u^{\mu}h_{\alpha}{}^{\nu}u^{\rho}h_{\beta}{}^{\sigma}$ and
$H_{\alpha\beta} :=
-\frac{1}{2}\epsilon_{\rho\sigma\gamma\delta}
C_{\mu\nu}{}^{\gamma\delta}u^{\rho}h_{\alpha}{}^{\sigma}
u^{\mu}h_{\beta}{}^{\nu}$ are the electric and magnetic Weyl
curvature tensors with respect to the matter frame.

The propagation equations for the scalar-valued matter variables 
read:\footnote{With Eq.~(\ref{def:kdef}),
when acting on scalar-valued variables, the
operator identity $E^{-1}k^{\rho}\partial_{\rho}
= u^{\rho}\partial_{\rho}- e^{\rho}\partial_{\rho}$ applies.}
\begin{subequations}
\bea
\lb{drthdv}
&&E^{-1}k^{\rho}\partial_{\rho}\th
=  -\frac{1}{3}\,\th^{2}-2\sig^{2}-4\pi\varrho
+\Lambda - \th ^{\prime} \ ;  \\
\lb{drhodv}
&&E^{-1}k^{\rho}\partial_{\rho}\varrho
=  -\th\varrho  -  \varrho^{\prime} \ ; \\
\lb{dsigmadv}
&&E^{-1}k^{\rho}\partial_{\rho}\sig^{2}
=  - \frac{4}{3}\,\th \sig^2
- \sig_{\mu}{}^{\nu}\sig_{\nu}{}^{\alpha}\sig_{\alpha}{}^{\mu} 
- (\sig^{\mu \nu}E_{\mu \nu}) - (\sig^2)^{\prime} ; \\
\lb{dprojsigmadv}
&&E^{-1}k^{\rho}\partial_{\rho}(p^{\mu \nu}\sig_{\mu \nu})
 =  - \frac{2}{3}\,\th (p^{\mu \nu}\sig_{\mu \nu})
-  p^{\mu \nu}\sig_{\mu \alpha}\sig_{\nu}{}^{\alpha}\nonumber\\
 &&\qquad\qquad\qquad\qquad\quad\ + \frac{4}{3}\,\sig^2
- (p^{\mu \nu}E_{\mu \nu}) 
- 2\sig_{\mu \nu} e^\mu \kappa^\nu
- (p^{\mu \nu}\sig_{\mu \nu})^{\prime} \ , 
\eea
\end{subequations}
where the operator ${}^{\prime} :=  e^\mu \nabla_\mu$ denotes the
spatial derivative along $e^\mu$, and
$(p^{\mu\nu}E_{\mu\nu})$ is the longitudinal projection of the
electric Weyl curvature tensor, which accounts for tidal effects
in the evolution equations.

We can recast Eqs.~(\ref{drthdv}) and (\ref{dprojsigmadv}) into
the following evolution equations for the area-adapted
longitudinal expansion rate variables $\alpha$ and $\beta$:
\begin{subequations}
\bea
\lb{dalphadv}
E^{-1}k^{\rho}\partial_{\rho}\alpha
& = & -\alpha^{2} - \alpha^{\prime} + (p^{\mu\nu}E_{\mu\nu})
- \frac{4\pi}{3}\varrho + \frac{1}{3}\Lambda
- (e^{\mu}\sigma_{\mu\alpha})(e^{\nu}\sigma_{\nu\beta})
p^{\alpha\beta} \nonumber \\
& & + \ 2\sigma_{\mu\nu}e^{\mu}\kappa^{\nu} \ ; \\
\lb{dbetadv}
E^{-1}k^{\rho}\partial_{\rho}\beta
& = & -\frac{1}{9}\,(\alpha+2\beta)(4\beta-\alpha)
- \beta^{\prime} - \frac{1}{2}\,(p^{\mu\nu}E_{\mu\nu})
- \frac{4\pi}{3}\varrho + \frac{1}{3}\Lambda \nonumber \\
& & - \ \frac{1}{2}\,(e^{\mu}\sigma_{\mu\alpha})
(e^{\nu}\sigma_{\nu\beta})p^{\alpha\beta}
- \frac{1}{2}\,\sigma_{\mu\alpha}\sigma_{\nu\beta}
p^{\mu\nu}p^{\alpha\beta}
- \sigma_{\mu\nu}e^{\mu}\kappa^{\nu} \ .
\eea
\end{subequations}

We may, in addition, formulate the evolution equation 
for the cosmological redshift, defined in Eq.~(\ref{redshift}).
Using Eq.~(\ref{opticalenergy}), we have
(\textit{cf.} Ref.~\ct[p.~147f]{ell1971})
\be
\lb{vtoztransf}
k^{\rho}\partial_{\rho}z = -\,E(P) \alpha (1+z)^{2} \ .
\ee
The cosmological redshift is a monotonic function along
the geodesic null curves only when the area-adapted longitudinal
expansion rate variable $\alpha$ does not change sign. 
Thus, as pointed out by R\"{a}s\"{a}nen
\ct{lightpropndust}, cosmological redshift can in general not be
used as a parameter along the geodesic null congruence when
scales of collapsing structures are considered; see
section~\ref{sec:discussion} for a discussion on valid
parametrizations of light fronts. For the same reason, cosmological
redshift cannot in general be thought of as a cosmological time
variable, since, typically, $u^{\rho}\partial_{\rho}z$ will change
sign along the world lines of the matter congruence. The proper
time function $\tau$ for the matter congruence, on the other hand,
is a monotonic function along the geodesic null congruence,
\be
\lb{tauparam}
k^{\rho}\partial_{\rho}\tau = -k^\mu u_\mu = E  \ , 
\ee
since $E$ is a positive-valued variable. We remark that level
surfaces $\{\lambda = \text{constant}\}$
-- with $\lambda$ defined as an affine parameter along the
geodesic null congruence satisfying
$k^{\rho}\partial_{\rho}\lambda = 1$
-- do not in general coincide with spacelike $3$-surfaces
$\{\tau = \text{constant}\}$. $E$ is in general an inhomogeneous
function on the screen space, which renders $\lambda$ an invalid
label for the light fronts $\{V = \text{constant},
\tau = \text{constant}\}$. 

\subsection{Constraint equations on the light fronts}
\lb{sec:constraints}
The null kinematic variables 
given in Eqs.~(\ref{kinhat}) and~(\ref{kintilde}) are subject to 
constraints from the embedding of the $2$-dimensional light fronts
into the cosmological spacetime;
\textit{cf.} Refs.~\ct{steetal2003}, \ct[p.~258]{wal1984}
and \ct[Sec.~16.2]{ste1991}. 
When expressed in terms
of the covariantly $(1+1+2)$-decomposed variables, the Gauss
embedding constraint reads:
\be
\lb{lightfrontgauss}
0 = K - (p^{\mu\nu}E_{\mu\nu})
- \frac{8\pi}{3}\varrho - \frac{1}{3}\Lambda  
+ \frac{1}{4}\,\frac{\hth}{E}\,\frac{\tith}{E}
- \frac{1}{2}\,\frac{\hsig_{\mu\nu}}{E}\,\frac{\tisig^{\mu\nu}}{E}
\ ,
\ee
while the two Codazzi embedding constraints are given by 
\begin{subequations}
\bea
\lb{lightfrontcodazziin}
0 & = & p^{\mu\alpha}p^{\nu\beta}
\D_{\nu}\left(\frac{\hsig_{\alpha\beta}}{E}\right)
- \frac{1}{2}\,p^{\mu\nu}\D_{\nu}\left(\frac{\hth}{E}\right)
+ \left(\frac{\hsig^{\mu\nu}}{E}
-\frac{1}{2}\,\frac{\hth}{E}\,p^{\mu\nu}\right)
\frac{\D_{\nu}E}{E} \ ; \\
\lb{lightfrontcodazziout}
0 & = & p^{\mu\alpha}p^{\nu\beta}
\D_{\nu}\left(\frac{\tisig_{\alpha\beta}}{E}\right)
- \frac{1}{2}\,p^{\mu\nu}\D_{\nu}\left(\frac{\tith}{E}\right)
+ \left(\frac{\tisig^{\mu\nu}}{E}
-\frac{1}{2}\,\frac{\tith}{E}\,p^{\mu\nu}\right)
\frac{\D_{\nu}E}{E} \ ,
\eea
\end{subequations}
respectively. In Eq.~(\ref{lightfrontgauss}) the extrinsic
curvature of the embedding has been formulated using the
area expansion tensors associated with $k^{\mu}$ and $l^{\mu}$.
The scalar $K := \frac{1}{2}{}^{(2)}\!R$ is the
Gaussian curvature of the screen space, 
and ${}^{(2)}\!R$ is the Ricci scalar of the same screen space. The
sectional curvature of the light fronts is given by the projection 
\be
\lb{seccurv}
\frac{1}{2}\,R_{\mu \nu \alpha \beta} \, p^{\mu\alpha}
p^{\nu \beta} = (p^{\mu \nu}E_{\mu \nu}) + \frac{8\pi}{3}\varrho
+ \frac{1}{3}\Lambda \ , 
\ee
where $R_{\mu \nu \alpha \beta}$ is the Riemann curvature tensor
for the cosmological spacetime. 


\subsection{Photon conservation, distance measures, and redshift
drift}
\label{sec:conservation}
In the present analysis we consider light rays propagating
along the observer's past null cone
${\cal C}^{-}(P):\!\{V = \text{constant}\}$
to be associated with freely streaming test particles which are
-- apart from the point of emission and the point of observation
at the vertex $P$ of the past null cone -- non-interacting with
the matter content of the cosmological spacetime. In this setting
we are entitled to assume the conservation of the current density,
\be
J^\mu = {\mathfrak n}\,k^{\mu} \ ,
\ee
where $\mathfrak{n}$ is the number density of light front
continuum elements (photons) with respect to the area enclosed by
the screen space. The conservation of $J^\mu$ leads to 
\be
\lb{photonconservation}
\nabla_\mu J^\mu =0 \; \Leftrightarrow \; 
k^{\rho}\partial_{\rho}{\mathfrak n}
= - \hth \,{\mathfrak n} \ ,
\ee 
i.e. the number of radiation elements per area element is
conserved. This conservation law is fundamental for the definition
of cosmological distance measures, see, e.g., Refs. ~\ct{ell1971}
and \ct{sasaki1}, and is an important assumption in order to
obtain the relativistically corrected luminosity distances. 

The \textit{observer area distance} $d_{A}$, henceforth referred to
as the \textit{effective angular diameter distance},\footnote{In
generic cosmological spacetimes the observer area distance for a
luminous astrophysical source does not necessarily coincide with
the linear size based angular diameter distance for the same
source due to potential Weyl curvature induced shearing along the
observer's past null cone of an incoming geodesic null ray bundle.
Only in an exact FLRW cosmology do their
conceptions become identical as a consequence of the prevailing
spatial isotropy, and thus vanishing Weyl curvature;
\textit{cf.} Refs.~\ct[Eq.~(25,27)]{ste1991} and
\ct[Sec.~4.5.2]{ellhve1999}.} for a luminous astrophysical source
is defined by \ct[Eq.~(6.27)]{ell1971}
\be
d_{A}^{2}  :=  \frac{{\delta}A}{{\delta}\Om} \ ,
\ee
where ${\delta}A$ is the physical area covered by the source
at its own position in directions perpendicular to the spatial
propagation direction of the emanating geodesic null rays,
and ${\delta}\Om$ is the solid angle the source subtends at the
position $P$ of the observer (with $0 < {\delta}\Om < 4\pi$).
The evolution of the effective angular diameter distance
along the geodesic null congruence $k^{\mu}$ is given by
(\textit{cf.} Ref.~\ct[Eq.~(6.19)]{ell1971})
\be
\lb{areadistevol}
E^{-1}k^{\rho}\partial_{\rho}d_{A} =
\frac{1}{2}\,\frac{\hth}{E}\,d_{A} \ ;
\qquad
E^{-1}k^{\rho}\partial_{\rho}d^{-2}_{A} =
-\,\frac{\hth}{E}\,d^{-2}_{A}\ .
\ee 
This, when combined with
Eq.~(\ref{opticalraychaudhuri}), yields the focusing equation
(\textit{cf.} Ref.~\ct[Eq.~(44)]{per2004}):
\be
\lb{focuseq}
E^{-1}k^{\rho}\partial_{\rho}(E^{-1}k^{\sigma}\partial_{\sigma}
d_{A})
= \left(\frac{1}{2}\,\alpha\,\frac{\hth}{E}
- \left(\frac{\hsig}{E}\right)^{2}-4\pi\varrho\right)d_{A} \ .
\ee
Using the conservation law (\ref{photonconservation}) and 
Etherington's reciprocity theorem \ct{Etherington,ellhve1999,Nez},
the luminosity distance between
a source and the observer at the vertex point $P$,
\be
\lb{luminosityd}
d_L  :=  \sqrt{\frac{L^\mathrm{bol}}{4\pi I^\mathrm{bol}}} \ ,
\ee
can be written in terms of the effective angular diameter distance
as follows: 
\be
\lb{luminosityd2}
d_L = (1+z)^2 d_A \ ;
\ee
$L^\mathrm{bol}$ is the bolometric luminosity of the
source, and $I^\mathrm{bol}$ is the bolometric flux of energy as
measured by the observer. 

Let us finally consider the drift of cosmological redshift of a
given source in time as measured by an observer at
the vertex point $P$. 
Let the source and the observer be comoving with the 
irrotational $4$-velocity field $u^\mu$ of the matter
congruence and let the incoming geodesic null rays connecting them
be generated by~$k^\mu$. The spacetime point of observation $P$ and
the spacetime point of the source $S$ are shifted along the matter
congruence at their respective locations, creating a change
in cosmological redshift which we measure in the observer's proper
time $\tau_0$ \cite{Heinesen:2020pms}:
\bea
\lb{zdrift}
\hspace*{-0.6cm} \frac{d z }{d\tau_0}
& = & u^{\rho}\partial_{\rho}z\,\lvert_{P}
+ \frac{d\tau}{d\tau_0}\,u^{\rho}\partial_{\rho}z\,\lvert_{S}
\, = \, (1+z)\,\alpha\,\rvert_{P}
- (1+z)\,\frac{E^\prime}{E}\,\bigg\rvert_{P}
- \alpha\,\rvert_{S}
+ \frac{E^\prime}{E}\,\bigg\rvert_{S}  ,
\eea
where $d\tau/d\tau_0 = 1/(1+z)$ is the Jacobian of the change
of the proper time measure between the source and observer
world lines from the bijection induced by the geodesic null
congruence. The second equality can be obtained by substituting
$u^\mu = k^\mu/E + e^\mu$ and using Eq.~(\ref{opticalenergy}).
We see that the cosmological redshift drift (\ref{zdrift})
depends on the area-adapted longitudinal expansion rate variable,
$\alpha$, of the matter congruence (replacing the Hubble parameter in the analogous FLRW expression for redshift drift) and on the spatial gradient of the photon energy, $E$, along the spatial direction of the incoming null ray. 

\section{Area-averaging over light fronts} 
We now consider the evolution of area-averaged physical
quantities along the observer's past null cone. 
In section \ref{sec:avevolution} we define the area-averaging
operation on the light fronts  
and construct evolution equations for the area of
the light fronts. 
In section \ref{sec:GB} we consider conservation
laws and the Gauss--Bonnet theorem as constraints for the average
dynamics of the light fronts.
In section \ref{sec:averageobs} we derive evolution equations for
area-averaged cosmological redshift and distance measures,
and in section \ref{sec:FLRW} we check the relations derived
against the FLRW limit.  

\subsection{Area-averaging operation and evolution equations for
area-averaged variables}
\lb{sec:avevolution}
We shall now define the covariant area-averaging operation on the
light fronts. 
The area-averaging operation defined is
invariant under coordinate transformations (though we shall define
useful area-adapted coordinates on the $2$-surfaces of
integration), and it can be formulated in terms of
covariant integrals in the cosmological spacetime, 
as detailed in Ref.~\cite{lightconeav1}.  

Consider an orientable and simply connected compact
$2$-dimensional domain $\CF$ in the light front
which lies within a $2$-surface of intersection between
the observer's past null cone ${\cal C}^{-}(P)$ and a spacelike
$3$-surface ${\cal S}(\tau)$, the latter labelled by the observer's
corresponding value of proper time: ${\CF}(\tau) \subseteq
({\cal C}^{-}(P) \cap {\cal S})(\tau)$. The total domain
$({\cal C}^{-}(P) \cap
{\cal S})(\tau)$ is assumed to be topologically closed (as, e.g.,
the all-sky last scattering surface of the Cosmic Microwave
Background radiation). For the general spatially inhomogeneous 
case, this assumption corresponds to a choice of closed space
form, e.g., a spherical topology in the simplest case. 
We shall allow ${\CF}(\tau)$ to be any compact $2$-dimensional
subdomain of $({\cal C}^{-}(P) \cap
{\cal S})(\tau)$ with boundaries propagated along $k^\mu$ such that we
track a system of constant number of radiation elements. 

\begin{figure}[ht!]	
\centering
\includegraphics[width=0.9\linewidth]{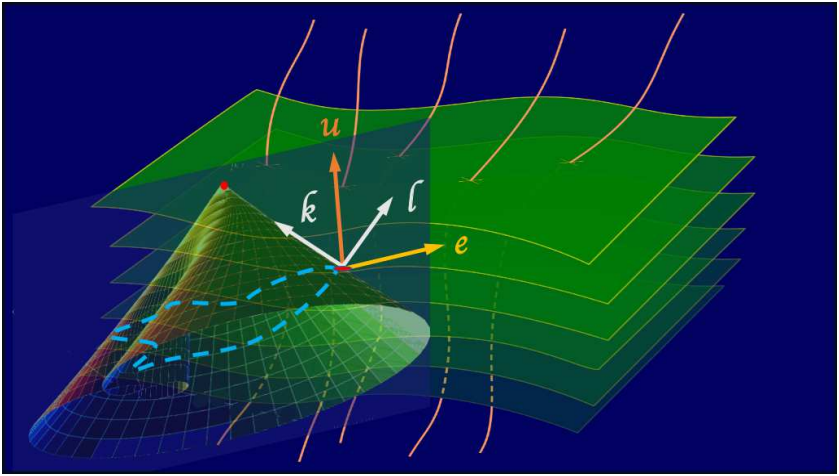}	
\caption{The $4$-vector fields
$\boldsymbol{k}$, $\boldsymbol{l}$, $\boldsymbol{u}$ and
$\boldsymbol{e}$ (right column of \eqref{vectorsincomponents})
at a line of intersection $\{V = \text{constant},
\tau = \text{constant}\}$ of the past null cone ${\cal C}^{-}(P)$,
embedded into the cosmological spacetime with GLC coordinates
$\{\tau, V, x^{A=2}\}$, and a single spacelike
$3$-surface ${\cal S}(\tau)$. The third dimension of the family of
spacelike $3$-surfaces $\{\tau = \text{constant}\}$,
as parametrized by $x^{A=3}$, is suppressed. The wave vector $\boldsymbol{k}$ is
tangent to the past null cone, while the wave vector
$\boldsymbol{l}$ represents an auxiliary null-like field pointing in the opposite spatial direction of $\boldsymbol{k}$ if 
seen in the restframe of the $4$-velocity $\boldsymbol{u}$, which is orthogonal to
the spacelike $3$-surfaces (containing the vector
$\boldsymbol{e}$ as a tangent). $\boldsymbol{u}$ is also, together with the
wave vector $\boldsymbol{k}$, orthogonal to the
$2$-dimensional light fronts $\{V = \text{constant},
\tau = \text{constant}\}$ (blue dashed line).
Recall the vector relations
\eqref{def:kdef} and \eqref{def:ldef}, and see related figures
in Ref.~\cite{Fleuryetal:GLC}. (Past null cone in this figure
inspired by Fig.~12 of Ref.~\cite{per2004}; spacetime foliation
in this figure: Pierre Mourier (priv. comm.).)}
\label{fig1}
\end{figure}

For the present purpose, it is convenient to use local
spacetime coordinates {$x^\mu = (\tau,V, x^A)$,} where $x^{A}$,
with indices $A = 2,3$, are local coordinates adapted to the
screen spaces such that they are constant along the geodesic 
null rays:\footnote{In general, local coordinates preserved
along $k^\mu$ are not preserved along $u^\mu$, since
$k^{\rho}\partial_{\rho}x^A = 0$ implies that
$k^\alpha \nabla_\alpha (u^{\rho}\partial_{\rho}x^A)
= \pounds_{\textbf{k}} (u^\rho) \partial_{\rho}x^A $, where the
operator $\pounds_{\textbf{k}}$ is the Lie derivative along the
geodesic null rays associated with $k^\mu$. Thus, if the
change of $u^\mu$ along the geodesic null rays has
components tangential to the screen space (which happens
generically in spatially inhomogeneous cosmologies),
then $u^{\rho}\partial_{\rho}x^A = 0$ cannot be satisfied globally on
the past null cone.}
$k^{\rho}\partial_{\rho}x^A = 0$.  
Gasperini, Veneziano and their collaborators introduced these
coordinates as \sayy{Geodesic
Light cone Coordinates (GLC)} with $\tau$ and $V$ exchanged; see Refs.~\cite{lightconeav1}, \cite{Fleuryetal:GLC} and
appendix~\ref{appA} for details, and [chap.3 and references
therein]\cite{elmardi:phd} for the relation to George Ellis'
observational \sayy{Perturbed Light cone Gauge} (PLG) coordinates.
In GLC coordinates the components of the cosmological
spacetime metric and its inverse read:
\be
\label{GLCmetric}
g_{\mu\nu} =
\left(
\begin{array}{ccc}
0 & 1/E & \vec{0} \\
1/E & \, (1/E^2) \! + \! U^2 \, & U_B \\
\vec{0} & U_A & p_{AB}  \\
\end{array}
\right) \ ;~~ ~~~~~
g^{\mu\nu} =
\left(
\begin{array}{ccc}
-1 & \, E \,  & - E U^B \\
E & 0 & \vec{0}  \\
- E U^{A} & \vec{0} & p^{AB} \\
\end{array}
\right) \ ,
\ee
and the screen space-forming vectors and associated
one-forms read (see Fig.~\ref{fig1}): 
\begin{align}
\lb{vectorsincomponents}
&k_\mu = (0, 1, 0, 0) \ ;  &&k^\mu = (E, 0, 0, 0) \ ;
\nonumber\\ 
&l_{\mu} = (-2E,-1,0,0) \ ; &&l^{\mu} = (E,-2E^2, 2E^2 U^A )
\ ; \nonumber\\
&u_\mu = (-1,0,0,0) \quad  ; &&u^\mu = (1, - E, EU^A) \ ;
\nonumber\\
&e_{\mu} = (-1,-1/E,0,0)\ ; &&e^{\mu} = (0,-E,EU^A)  \ ,
\end{align}
with the coordinate-independent projections $k_\mu k^\mu = l_\mu l^\mu = u^\mu e_\mu = u_\mu e^\mu= 0$,
$u_\mu u^\mu = -1$, $e_\mu e^\mu = 1$,
$k_\mu u^\mu = k^\mu u_\mu = l_\mu u^\mu = l^\mu u_\mu
= k_\mu e^\mu = k^\mu e_\mu = - E$, $l_\mu e^\mu = l^\mu e_\mu
= E$, $k_\mu l^\mu = k^\mu l_\mu = - 2E^2$. 
The $2$-dimensional tensor $p_{AB}$ is the induced metric
on the screen space and $p^{AB}$ is its inverse, while the
$2$-dimensional vector $U^A = u^\rho \partial_\rho (x^A) /E$
defines the drift of the area-adapted coordinates
$x^{A}$ along the matter congruence, and $U^2 := p_{AB} U^A U^B$.
\\
We define the area of a $2$-surface within a screen space by 
\be
A_{\CF}(\tau) = | \CF(\tau) |
= \int_{\CF}\sqrt{p(\tau,x^{A})}\,{\rm d}^{2}x \ ,
\ee
where $p(\tau,x^{A})
:= \det(p_{AB})$ is the Riemannian area element of
the $2$-surface. We define the area-average of a general scalar-valued
function $F(\tau,x^{A})$ as follows:
\be
\la F\ra_{\CF}\,(\tau)  :=  \frac{1}{A_{\CF}(\tau)}
\int_{\CF}F(\tau,x^{A})\sqrt{p(\tau,x^{A})}\,{\rm d}^{2}x \ .
\ee
It shall also be convenient to define area-averaging of
products of scalar-valued functions $F$ and $G$ as evaluated at
different points along the geodesic null rays (parametrized
in terms of the observer's proper time function $\tau$):
\bea
\lb{productav}
&& \la F(\tau,x^{A}) G(\tau',x^{A}) \ra_{\CF}  :=
\nonumber \\
&& \frac{1}{A_{\CF}(\tau)}
\int_{\CF}F(\tau,x^{A}) G(\tau',x^{A})  \sqrt{p(\tau,x^{A})}\,
{\rm d}^{2}x \ , 
\eea
where the evaluation of the area element is at the unprimed
coordinate time $\tau$.

The commutation rule for the operations of area-averaging and
evolution along the observer's past null cone is 
\bea
\lb{com}
\ptl_{\tau}\la F\ra_{\CF}
- \La \ptl_{\tau} F \Ra_{\CF}
& = &  \text{Cov}_{\CF}\left(\frac{\hth}{E}\, , \,F\right) \ ,
\eea
where $ \ptl_{\tau} := E^{-1}k^{\rho}\partial_{\rho}$ is the directional derivative along the null congruence as measured in units of $\tau$ (cf. Eq.~(\ref{tauparam}) for the Jacobian,
$k^{\rho}\partial_{\rho} \tau = E$), and where, for scalar-valued functions $X$ and $Y$:
\bea
\lb{covariance}
\text{Cov}_{\CF}(X , Y)
 :=  \la X Y  \ra_\CF  - \la X \ra_\CF \la Y  \ra_\CF \ . 
\eea
The local evolution of a scalar-valued function per unit proper
time is naturally measured along the geodesic null rays
of $k^\mu$, since the boundaries of the spatial domain as well as
the past null cone scalar $V$ are comoving with $k^\mu$. 
In deriving Eq.~(\ref{com}), the Jacobi identity,
\be
\lb{jacid}
E^{-1}k^{\rho}\partial_{\rho}\sqrt{p(\tau,x^{A})}
= \ptl_{\tau} \sqrt{p(\tau,x^{A})}
= \frac{\hth}{E}\,\sqrt{p(\tau,x^{A})} \ ,
\ee
has been used; \textit{cf.} Refs.~\ct[Eq.~(6.19)]{ell1971} and
\ct[Eq.~(3.14)]{sasaki1}. 
We define the domain-dependent dimensionless area scale factor
for the light fronts:
\be
\lb{areacalefactor}
a_{\CF}(\tau)  := 
\left(\frac{A_{\CF}(\tau)}{A_{{\CF}_{i}}}\right)^{1/2} \ , 
\ee
where the reference area $A_{{\CF}_{i}} := A_{\CF}(\tau_i)$ is
evaluated at a reference light front domain  
with time label $\tau_i$. Since the domain ${\CF}(\tau)$ is defined as
having no flow of geodesic null rays across its boundary, it fails
in general to preserve the number of matter fluid elements. Thus,
$a_{\cal F}(\tau)$ \textit{cannot} be interpreted as an
average scaling of the coordinate distance between individual
world lines of the matter congruence, and so does
\textit{not} reduce to the FLRW scale factor in the case of
spatial homogeneity and isotropy on the spacelike
$3$-surfaces orthogonal to the matter congruence.  
We formulate the evolution along the observer's past null
cone ${\cal C}^{-}(P):\!\{V=\text{constant}\}$ of the area scale
factor as a first part of a theorem. 
\begin{thgroup}
\label{ths:av}
\begin{theorem}[Effective evolution equations for light fronts]
\label{th:av_1} \\
The evolution along the observer's past null cone
${\cal C}^{-}(P):\!\{V=\text{constant}\}$ of the area scale
factor is governed by the \textit{screen space area expansion
rate}:
\be
\lb{areaexpansion}
\mathcal{H}_{\CF}
:= \frac{\ptl_{\tau}a_{\CF}(\tau)}{a_{\CF}(\tau)}
= \frac{1}{2} \La \frac{\hth}{E} \Ra_{\CF} \ ,
\ee
where we defined the \textit{$2$-surface area expansion
functional} $\mathcal{H}_{\CF}$.
The second derivative of the area scale factor yields the
{\rm `screen space area acceleration law'}:
\be
\lb{areaacceleration}
\frac{\ptl^2_{\tau}a_{\CF}(\tau)}{a_{\CF}(\tau)} 
= - 4 \pi  \La \varrho  \Ra_{\CF} 
+ \mathcal{Q}^{\mathbf k}_{\CF}  + {\mathcal S}^{\mathbf k}_{\CF}
\ ,   
\ee
where we have used the optical evolution equations
(\ref{opticalenergy}) and (\ref{opticalraychaudhuri}). The
function
\bea
\lb{Q1}
\mathcal{Q}^{\mathbf k}_{\CF} & := &
\frac{1}{4}\,\text{\rm Cov}_{\CF}\left(\frac{\hth}{E} \, , \,
\frac{\hth}{E}  \right)  -  \La \frac{\hsig^2}{E^2}\Ra_{\CF}  
\eea
is a {\rm `screen space kinematic backreaction'} term arising
from local spatial inhomogeneity and anisotropy of the incoming
geodesic null ray bundle associated with~$k^{\mu}$.
The function
\be
\label{lapse}
{\mathcal S}^{\mathbf k}_\CF := \frac{1}{2}
\La  \frac{\hth}{E}\,\alpha \Ra_{\CF}
= \frac{1}{2} \La  \frac{\hth}{E}\,k^{\rho}\partial_{\rho}
\left(\frac{1}{E}\right)\Ra_{\CF}
= -\frac{1}{2} \La  \frac{\hth}{E}\,
\frac{\ptl_{\tau} E}{E}\Ra_{\CF}
\ee
arises from the re-parametrization of the incoming
geodesic null congruence and measures the failure of $\tau$ to
be an affine parameter along this null congruence.
\end{theorem}
\noindent
\textbf{Remarks to Theorem \ref{th:av_1}}

Removing the term \eqref{lapse} is possible by averaging over
$\{\lambda = \text{constant}\}$ light fronts instead of
$\{\tau = \text{constant}\}$ light fronts. We stick to the
$\{\tau = \text{constant}\}$ foliation of the observer's past null
cone in the present paper. The term ${\mathcal S}^{\mathbf k}_\CF$
vanishes for energy functions
$E = k^{\rho}\partial_{\rho}\tau$, \textit{cf.}
Eq.~\eqref{tauparam}, that are constant along the integral curves
of the geodesic null congruence, i.e., when the geodesic
null rays are not subject to cosmological redshift.
The function $1/E$ plays the role of an inhomogeneous
lapse function along the geodesic null congruence,
\textit{cf.} Eq.~\eqref{GLCmetric} and
appendix~\ref{appA}.\footnote{For a discussion of the degeneracy
at the null cone of the standard ADM (Arnowitt, Deser and Misner)
$3+1$ slicing formalism \cite{ADM} and its generalization for
lightlike foliations, exemplified for double null foliations,
see Refs.~\cite{Israel:nullADM} and~\cite{Israel:doublenull}.}

\medskip
We formulate the area-average of the Gauss embedding
constraint~(\ref{lightfrontgauss}) as a second part of this 
theorem. 
\begin{theorem}[Effective energy constraint on light fronts]
\label{th:av_2} \\
Averaging the Gauss embedding
constraint~(\ref{lightfrontgauss}) over the compact domain~$\CF$
of the light fronts yields the \rm{`screen space area expansion law'}:
\bea
\lb{avconstraint2}
\mathcal{H}_{\CF}^2
+ \mathcal{H}_{\CF} \La  \overline{\theta}  \Ra_{\CF}
& = &  \frac{8 \pi}{3} \La \varrho  \Ra_{\CF}
+ \frac{1}{3}\Lambda - \La K  \Ra_{\CF}
-  \mathcal{Q}^{\mathbf{k}\mathbf{l}}_{\CF} \ , 
\eea
where
\bea
\lb{X1}
&&\mathcal{Q}^{\mathbf{k}\mathbf{l}}_{\CF}  :=  
\frac{1}{4}\,\text{\rm Cov}_{\CF}\left(\frac{\hth}{E} \, , \,
\frac{\tith}{E} \right)
- \frac{1}{2}
\La \frac{\hsig_{\mu\nu}}{E}\,\frac{\tisig^{\mu\nu}}{E}\Ra_{\CF} 
- \La p^{\mu\nu}E_{\mu\nu} \Ra_{\CF}
\eea
is a second backreaction term on the screen space, which may
be substituted by the linearly transformed backreaction variable,
using the relations (\ref{inout}):
\bea
\lb{Y1}
&&\mathcal{Q}^{\mathbf{k}\mathbf{e}}_\CF :=
\mathcal{Q}^{\mathbf{k}\mathbf{l}}_{\CF}
- \mathcal{Q}^{\mathbf{k}}_\CF
= \frac{1}{2}\,\text{\rm Cov}_{\CF}\left(\frac{\hth}{E} \, , \,
\overline{\theta} \right)
- 2\La \frac{\hsig_{\mu\nu}}{E}\ \overline{\sigma}^{\mu\nu}
\Ra_{\CF} 
- \La p^{\mu \nu}E_{\mu \nu}  \Ra_{\CF} \ . 
\eea
\end{theorem}

\noindent
\textbf{Remarks to Theorem \ref{th:av_2}}

Notice that the new backreaction term (\ref{X1}) contains products
of the kinematic variables associated with \textit{both}
the incoming and outgoing null congruences, $k^{\mu}$
and $l^{\mu}$, contrary to the backreaction
term (\ref{Q1}).\footnote{We remark that this backreaction term
appears to bear a relationship to the Hawking--Hayward energies as
defined in Refs.~\cite{Hawking,Hayward}; see
Refs.~\cite{bengtsson:hawkingenergy,Dennis1,Dennis2} for recent
work.}
Moreover, in both backreaction terms, (\ref{X1}) and
(\ref{Y1}), tidal effects enter through the screen space projection
of the electric Weyl curvature tensor.  
This is contrary to the $3$-dimensional volume-averaging
operation of Refs.~\cite{buchert:grgdust,buchert:grgfluid}, where
Weyl curvature only enters implicitly 
in the large-scale volume evolution.

\medskip
Since Eq.~(\ref{avconstraint2}) must be the integral of
Eq.~(\ref{areaacceleration}), we take the derivative of
Eq.~(\ref{avconstraint2}) with respect to $\tau$ and re-insert 
Eqs.~(\ref{avconstraint2}) and (\ref{areaacceleration}) to obtain
the following \textit{integrability condition} on the light fronts
that we formulate as a third part of theorem \ref{ths:av}.
\begin{theorem}[Integrability condition]
\label{th:av_3} \\
A necessary condition of integrability for
Eq.~(\ref{areaacceleration}) to yield Eq.~(\ref{avconstraint2})
is given by the integral constraint:
\begin{subequations}
\lb{integrability}
\bea
\lb{integrabilityexplicit}
&& \ptl_{\tau}  \mathcal{Q}^{\mathbf{k}}_{\CF}
+ 4 \mathcal{H}_{\CF} \mathcal{Q}^{\mathbf{k}}_{\CF}
+ \ptl_{\tau} \mathcal{Q}^{\mathbf{k}\mathbf{e}}_\CF
+ 2 \mathcal{H}_{\CF} \mathcal{Q}^{\mathbf{k}\mathbf{e}}_{\CF}
+ \ptl_{\tau} \La K \Ra_{\CF}
+  2 \mathcal{H}_{\CF} \La K \Ra_{\CF}  \nonumber  \\ 
&+& \ptl_{\tau} \left(  \mathcal{H}_{\CF}  
\La \overline{\theta} \Ra_{\CF}  \right)
+ 2 \mathcal{H}_{\CF} 
\left( \mathcal{H}_{\CF} \La \overline{\theta} \Ra_{\CF}\right)
+ 2 \mathcal{H}_{\CF} \left( {\mathcal S}^{\mathbf k}_\CF
- \frac{1}{3} \Lambda \right) \nonumber\\
& = &  \frac{8 \pi}{3} \left( \ptl_{\tau} \La \varrho  \Ra_{\CF}
+ 5 \mathcal{H}_{\CF}  \La \varrho  \Ra_{\CF} \right) \ , 
\eea  
where it follows from averaging of the local law for rest mass
flow Eq.~(\ref{drhodv}) that 
\bea
\lb{rhoav}
&&\ptl_{\tau} \La \varrho  \Ra_{\CF}
+ \La \theta  \Ra_{\CF}  \La \varrho  \Ra_{\CF}
=  -  \la  \varrho^{\prime} \ra_{\CF} 
- \text{\rm Cov}_{\CF}\left(\theta \, , \, \varrho  \right)
+  \text{\rm Cov}_{\CF}\left(\frac{\hth}{E} \, , \, \varrho \right)
\ .  
\eea
The integrability condition in Eq.~(\ref{integrabilityexplicit})
can be written in compact form:
\bea
\lb{integrabilitycompact}
&&\frac{1}{a_\CF^4} \ptl_\tau \left[ \mathcal{Q}^{\mathbf{k}}_\CF
a_\CF^4 \right] + \frac{1}{a_\CF^2} \ptl_\tau
\left[ \left( \mathcal{Q}^{\mathbf{k}\mathbf{e}}_{\CF}
+ {\mathcal S}^{\mathbf k}_\CF - \frac{1}{3} \Lambda
+ \La K \Ra_{\CF}
+ \mathcal{H}_{\CF} \La \overline{\theta} \Ra_{\CF}\right)
a_\CF^2 \right] \nonumber\\
& = & \frac{8 \pi}{3}\frac{1}{a_\CF^5} \ptl_\tau
\left[ \La \varrho \Ra_{\CF}  a_\CF^5 \right]
+ \ptl_\tau {\mathcal S}^{\mathbf k}_\CF \ .
\eea 
\end{subequations}
\end{theorem}
\end{thgroup}
\noindent
\textbf{Remarks to Theorem \ref{th:av_3}} 

Looking at Eq.~\eqref{integrabilitycompact}, we appreciate that
separate conservation of screen space kinematic backreaction
implies the following scaling behaviour
$\mathcal{Q}^{\mathbf{k}}_\CF
\propto a_\CF^{-4}$. Furthermore, separate conservation of the
area-averaged Gaussian curvature implies $\La K \Ra_{\CF} \propto
a_\CF^{-2}$, being inversely proportional to the square of the
area scale factor; see section~\ref{sec:GB} below for a discussion
on the Gauss--Bonnet theorem and area-averaged Gaussian curvature.
(Compare also Ref.~\cite{GBC} for a discussion of the averaged
equations in a $(2+1)$-spacetime.) The scaling behaviours derived
from requiring separate conservation laws of the cosmological
variables in Eq.~\eqref{integrabilitycompact} can in some cases
provide some guidance, but will in some cases also be misleading.
In general we expect non-trivial couplings between the various
macroscopic variables.\\

\medskip
\noindent
{\bf Discussion of Theorem \ref{ths:av}} 

The area-averaged equations (\ref{areaexpansion}),
(\ref{areaacceleration}), (\ref{avconstraint2}) and
(\ref{integrability}) appear more involved than
the corresponding evolution equations for the volume of
fluid-orthogonal spacelike $3$-surfaces in an irrotational
dust universe given in Ref.~\cite{buchert:grgdust}; \textit{cf.}
Ref.~\cite{buchert:grgfluid} for the
corresponding formulation including non-zero pressure in the fluid
description. This is not surprising, since the consideration of
generic cosmological dynamics and corresponding observations along
the observer's past null cone introduces additional variables. In
particular, in a covariant $(1+1+2)$-decomposition we see the
effects induced by the Weyl curvature. Even though the incoming
geodesic null rays in the present description are non-gravitating
test particles -- and thus do not collectively
act as a source in Einstein's field equations -- they enter in the
description of the observer's past null cone and in the
evolution along this past null cone of quantities to be measured. 

At this stage, the screen space area acceleration law
(\ref{areaacceleration}) of theorem \ref{ths:av} for the area scale
factor $a_\CF (\tau)$ in \ref{th:av_1} is subject to the
area-averaged constraint equation (\ref{avconstraint2}) in
\ref{th:av_2}. These equations involve three backreaction terms,
$\mathcal{Q}^{\mathbf{k}}_{\CF}$,
$\mathcal{Q}^{\mathbf{k}\mathbf{l}}_\CF$ and 
$\mathcal{Q}^{\mathbf{k}\mathbf{e}}_\CF$, where the latter two are
trivially related to the first, 
$\mathcal{Q}^{\mathbf{k}\mathbf{l}}_\CF
- \mathcal{Q}^{\mathbf{k}\mathbf{e}}_\CF
=  \mathcal{Q}^{\mathbf{k}}_{\CF}$. 
The integrability condition \eqref{integrability} in \ref{th:av_3}
serves as a balance equation among the various sources and
backreaction terms, and is not an independent equation.
In addition, these equations involve the further variables
$\La  \overline{\theta}  \Ra_{\CF}$, the area-averaged rest mass
density $\La \varrho \Ra_\CF$, the area-averaged Gaussian
curvature scalar $\La K \Ra_\CF$, and the
non-affine term ${\mathcal S}^{\mathbf k}_\CF$ \eqref{lapse}.
This forms a set of two independent equations for seven unknown
functions, in contrast to the $3$-dimensional dust case in
Ref.~\cite{buchert:grgdust}, where the counting yields 
three independent equations (those that correspond to the two
above, but also the continuity equation for the averaged fluid
density) for four unknown functions. However, the present case is
to be formally\footnote{We still refer to the dust case here; a
generalization including fluid pressure with general considerations
on lapse and shift can be found in \cite{fanizza:pressure}.}
compared with the 
$3$-dimensional case in Ref.~\cite{buchert:grgfluid}, where the
counting also yields two independent equations for seven unknown
functions (that can be reduced by one via an equation of state).
Unlike the dust case \cite{buchert:grgdust}, but similar to
the more involved case \cite{buchert:grgfluid}, the evolution
equations for $\ptl_{\tau}\La \overline{\theta} \Ra_{\CF}$ and
$\ptl_{\tau}\La \varrho  \Ra_{\CF}$, which can be derived from
Eqs.~\eqref{drthdv} and \eqref{drhodv}, i.e. $\ptl_{\tau}
\left( \mathcal{H}_{\CF}\La \overline{\theta}
\Ra_{\CF} \right) + 2 \mathcal{H}_{\CF} \left( \mathcal{H}_{\CF}\La
\overline{\theta}\Ra_{\CF} \right) = ...$, 
$\ptl_{\tau} \La \varrho  \Ra_{\CF}+ 5\mathcal{H}_{\CF} \La \varrho 
\Ra_{\CF} = ...$, 
do not immediately serve to constrain the system further due to
failure of the light front domain to be comoving with the matter
congruence, and due to the appearance of further  backreaction
terms. 

As detailed in the next section, invoking the Gauss--Bonnet
theorem has the potential to add a further constraint to the  
non-closed system of area-averaged equations.

\subsection{Average conservation laws and integral-geometric
measures} 
\lb{sec:GB}
Conservation laws can play an important role in the closure of
averaged evolution equations. In this section we investigate the
possibility of defining globally conserved quantities over the
light fronts. The local photon number conservation law
(\ref{photonconservation}) leads to the area-averaged equivalent: 
\be
\lb{avphotonconservation}
\ptl_{\tau}\la {\mathfrak n} \ra_{\CF}
+ 2 \mathcal{H}_{\CF}  \la {\mathfrak n} \ra_{\CF} = 0
\ \Leftrightarrow \  \la {\mathfrak n} \ra_{\CF}
=  \frac{\La {\mathfrak n} \Ra_{{\CF}_i}  }{a^2_{\CF}(\tau) } \ . 
\ee
In the case when we consider a global area-average over
a total light front domain, a $2$-surface $\Sigma :=
({\cal C}^{-}(P) \cap {\cal S})(\tau)$ which we assume to be
compact and topologically closed, we may invoke the Gauss--Bonnet
theorem for the area-averaged Gaussian curvature scalar $K$: 
\begin{subequations}
\be 
\lb{GBclosed}
\La K \Ra_{\Sigma}
= \frac{\La K \Ra_{{\Sigma}_i} }{a^2_{\Sigma}(\tau)} \ , \qquad
\La K \Ra_{{\Sigma}_i}  A_{{\Sigma}_i} = 2\pi \chi \ , 
\ee
where $\chi = 2-2g$ is the Euler characteristic of  
$\Sigma$ with genus $g$. 
Thus, in this case the area-averaged Gaussian curvature scalar
obeys its own separate conservation equation and decouples from
the integrability condition~(\ref{integrabilityexplicit}).
For light fronts
$\Sigma$ of spherical topology we have $\chi =2$, and so
$\La K \Ra_{{\Sigma}_i}  A_{{\Sigma}_i} = 4\pi$. 
(Level $2$-surfaces of a past null cone must always have
spherical topology, unless caustics change the topology by
destroying the bijection between the surfaces \cite{ellissolomons,mustapha3,Dennis1}.)

In the case of a light front \textit{with boundary}, we may
invoke the Gauss--Bonnet theorem with boundary term, now
considering a compact domain $\CF \subset \Sigma$: 
\be 
\lb{gaussbonnet}
\La K \Ra_{{\CF}}  A_{\CF} \, 
+ \int_{\partial\CF} \kappa \,{\mathrm d}{\ell}= 2\pi \chi \ ,
\ee
\end{subequations}
where $\kappa$ is the extrinsic curvature of the $1$-dimensional
boundary $\partial\CF$ of $\CF$, and
${\mathrm d}{\ell}$ is the line element for this boundary.  

We notice from the above considerations the natural appearance of 
integral-geometric measures of the light fronts, which can be
viewed as generalizations of the Minkowski functionals of convex
sets in Euclidian space; \textit{cf.} 
Ref.~\cite{beyond} and references therein: the $2$-surface area,
$A$ (content), the length of the circumference,
$L: = \int_{\partial\CF} {\mathrm d}{\ell}$ (shape), and the
Euler characteristic, $\chi$ (connectivity). 
As the area of a light front evolves, the shape of the
light front also evolves, while its Euler characteristic
is preserved according to our assumption of caustic-free
evolution along the observer's past null cone,
irrespective of the choice of compact submanifold of the
total light front domain $({\cal C}^{-}(P) \cap {\cal S})(\tau)$.

\medskip
The evolution of the
area-averaged Gaussian curvature scalar $\La K \Ra_{\CF}$ is in
general determined by the evolution  
of the area of the $2$-surface $A_{\CF}(\tau)$ and the
boundary term $\int_{\partial\CF} \kappa {\mathrm d}{\ell}$.
Whereas the evolution of the area is given by the area-averaged
expansion rate scalar of the geodesic null congruence, implicitly
given by Eq.~\eqref{areaexpansion}, the evolution of the boundary
term involves shear degrees of freedom as well as Weyl
curvature degrees of freedom. 

Let the boundary at each $2$-dimensional screen space be
determined by the level surfaces of value $B_0$ of a
spacetime function $B$, such that
\be
\lb{normalb}
n^\mu :=  \frac{p^{\mu \nu} \partial_\nu B }{ \mathcal{N}} \ ;
\qquad
\mathcal{N} := \sqrt{p^{\mu \nu}\partial_\mu B\,\partial_\nu B}
\ ; \qquad k^{\rho}\partial_{\rho}B = 0 \ ,
\ee
defines the outward directed normal to the
boundary.\footnote{For an analogous definition of domain
boundaries in the $3+1$ slicing formalism, see Ref.
\cite{Gasperini:2009mu}.} The last
condition ensures that the boundaries are constant along the
integral curves of the geodesic null congruence generating the
observer's past null cone. The normalized tangent vector of
the curve in the light front plane, $q^\mu$, is determined by:
\be
\lb{tangent}
q^\mu n_\mu = 0 \ ; \qquad q^\mu u_\mu = 0 \ ; \qquad
q^\mu e_\mu = 0 \ , 
\ee
and the induced metric on the boundary is given by
\be
\lb{tangent2}
q_\mu q^\nu= p_{\mu}{}^\nu - n_\mu n^\nu 
= \delta_\mu{}^\nu + u_\mu u^\nu - e_\mu e^\nu  - n_\mu n^\nu \ , 
\ee
such that $\{u^\mu, e^\mu, n^\mu, q^\mu \}$ constitute
an orthonormal vector basis for the cosmological
spacetime. The extrinsic curvature associated with the embedding
of the boundary in the $2$-dimensional screen space can now be
expressed as follows:
\be
\lb{kappa}
\kappa := n^\nu q^\mu \nabla_\mu q_\nu \ .
\ee
From the normalization requirement,
$n_\mu k^\nu  \nabla_\nu n^\mu = 0$, and the propagation rule,
$k^{\rho}\partial_{\rho}B = 0$, we have
\bea
\lb{evolN}
 \frac{k^{\rho}\partial_{\rho}\mathcal{N}}{\mathcal{N}}
= -  \frac{1}{2}\, \hth
- n^{\mu} n^\nu \hsig_{\mu \nu}\ .
\eea
The evolution of the normal $n^\mu$ from one screen space to
the next is given by
\bea
\lb{normalbev}
k^\nu \nabla_\nu n^\mu &=& - q^\mu \, q^{\alpha} n^\nu
\hsig_{\alpha  \nu} - k^\mu \, n^\alpha e^\nu \sig_{\alpha \nu} \ ,
\eea
which follows from the definition of $n^\mu$ in
Eq.~(\ref{normalb}). 
Similarly, we may compute the evolution of $q^\mu$ along the
geodesic null congruence:
\bea
\lb{tangentev}
k^\nu \nabla_\nu q^\mu &=& n^{\mu} \, q^{\alpha} n^\nu
\hsig_{\alpha  \nu} - k^\mu \, q^\alpha e^\nu \sig_{\alpha \nu} \ , 
\eea
which follows from the orthogonality requirements (\ref{tangent}). 
The evolution of the boundary term is then given by
\be 
\lb{boundaryevolution}
\ptl_{\tau} \int_{\partial\CF} \kappa\, {\mathrm d}{\ell}
= \int_{\partial\CF} \frac{ \nabla_\mu \left( \kappa\,
\mathcal{N} \, k^\mu \right) }{\mathcal{N} E} \,{\mathrm d}{\ell}
= \int_{\partial\CF}  \left(E^{-1}k^{\rho}\partial_{\rho}\kappa
+ \frac{\kappa\, q^{\mu} q^{\nu} \nabla_\mu k_\nu }{E}
\right) {\mathrm d}{\ell} \ , 
\ee
where Eqs.~(\ref{tangent}) and (\ref{evolN}) have been used, and
where
\be 
\lb{exppath}
q^{\mu} q^{\nu} \nabla_\mu k_\nu
=  \frac{1}{2} \hth + q^{\mu} q^\nu \hsig_{\mu \nu}  
\ee
is the projected expansion tensor along the tangent of the curve,
describing the expansion of the length of the boundary. 

An important observation is that the intermediate equality in
Eq.~(\ref{boundaryevolution}) shows that the boundary term is
conserved, if $(\kappa\, \mathcal{N} k^\mu)$ is a conserved
current in the vicinity of the boundary, corresponding to
$\kappa$ being a conserved density on the boundary domain. A
further insight is obtained when considering the evolution of
the extrinsic curvature of the boundary:
\bea
\lb{evkappa}
k^{\rho}\partial_{\rho}\kappa 
& = & - \left( \frac{1}{2}\,\hth
+ q^{\mu} q^\nu \hsig_{\mu \nu}\right)\left(\kappa-
n^\alpha e^\beta \sig_{\alpha \beta}\right)
- n^{\mu} q^\nu \hsig_{\mu \nu}\left(
q^\alpha e^\beta \sig_{\alpha \beta}
-\frac{q^{\rho}\partial_{\rho}E}{E}\right) \nonumber  \\ 
& + & q^{\alpha} \nabla_\alpha (n^{\mu} q^\nu \hsig_{\mu \nu})
- E\left(E_{\mu\nu}e^{\mu}n^{\nu}
+\eps_{\mu\nu}{}^{\beta}H_{\alpha\beta}n^{\mu}q^{\nu}q^{\alpha}
\right) \ .
\eea
In deriving Eq.~(\ref{evkappa}), the evolution equations
(\ref{normalbev}) and (\ref{tangentev}) have been invoked and it
has been used that
\bea
\lb{weylproj}
\hspace*{-0.2cm} n^\mu k^\nu q^\alpha q^\beta
R_{\mu \alpha \nu \beta}
= n^\mu  k^\nu q^\alpha q^\beta C_{\mu \alpha \nu \beta}
= - E\left(E_{\mu\nu}e^{\mu}n^{\nu}
+\eps_{\mu\nu}{}^{\beta}H_{\alpha\beta}n^{\mu}q^{\nu}q^{\alpha}
\right) , 
\eea
where the first equality follows from the form of the
energy-momentum tensor (\ref{eq:dust}), and the second equality
follows from the decomposition of the Weyl curvature tensor into
its electric and magnetic parts with respect to the
matter frame \cite{Maartens:1997fg}. 

It can be seen from Eq.~(\ref{evkappa}) that vanishing
on the boundary of each of the projected Weyl curvature terms of
Eq.~(\ref{weylproj}), the off-diagonal components of the projected
volume shear rate of the matter congruence,
$n^\alpha e^\beta \sig_{\alpha \beta}$, and the incoming optical
shear, $n^{\mu} q^\nu \hsig_{\mu \nu}$,\footnote{Any restriction
imposed on the volume shear rate of the matter congruence
leads to constraints on the Weyl curvature and on the spatial
gradient of the expansion rate through the geodesic deviation
equation and constraint equations for the relevant congruence
\cite{Ellis:2011}. For example, $n^{\mu} q^\nu \hsig_{\mu \nu} = 0
\Rightarrow k^\alpha  n^\mu q^\beta  k^\nu 
C_{\alpha ( \mu \beta ) \nu } = 0$ in the present set-up. By
imposing such restrictions it must be checked whether
there exist non-trivial cosmological spacetimes fulfilling such
conditions for the volume shear rate.} results in conservation of
the boundary term. In this special case, we have from
Eq.~(\ref{gaussbonnet}) that $\La K \Ra_{{\CF}} \propto 1/A_{\CF}$,
with the constant of proportionality depending on the initial
boundary chosen. This is a scaling behaviour of the
area-averaged Gaussian curvature scalar
equivalent to that of the boundary-free case. Given the strict
constraints which must be imposed in order to mimic the
boundary-free case, the scaling law $\La K \Ra_{{\CF}}
\propto 1/A_{\CF}$ is likely not satisfied in 
realistic cosmological spacetimes.

\subsection{Area-averaged observables}
\lb{sec:averageobs}
We shall now perform the area-average over observable quantities
for an observer viewing multiple objects on the same
$2$-dimensional screen space. Using Eqs.~(\ref{redshift}) and
(\ref{vtoztransf}), the evolution of the area-average of the
logarithmic cosmological redshift
along the observer's past null cone reads:
\be
\lb{aveq4}
\ptl_{\tau}\la \ln(1+z) \ra_{\CF}
= - \La \alpha \Ra_{\CF}
+ \text{Cov}_{\CF}\left(\frac{\hth}{E} \, , \, \ln(1+z) \right) \ . 
\ee 
Unlike the local cosmological redshift
function, which, in general, cannot be thought of as a
time variable along the world lines of the matter congruence,
the area-averaged logarithmic cosmological redshift can be used as
a time label for the $2$-dimensional light fronts as long as the
right-hand side of Eq.~(\ref{aveq4}) does not change sign. 

We can, furthermore, compute the evolution along the observer's
past null cone of the area-average of the logarithm of the
effective angular diameter distance, which, by
Eq.~(\ref{areadistevol}), reads:\footnote{We omit the
normalization of the dimensionfull quantities $d_A$ and $d_L$
inside the logarithm; the corresponding equations hold for any
choice of normalization.}
\bea
\lb{avareadistevol}
&& \ptl_{\tau}\la \ln(d_A) \ra_{\CF} 
= \mathcal{H}_{\CF} + \int^{\tau}
\mathcal{M}(\tau,\tau^{\prime})\,{\mathrm d}\tau^{\prime} \ , 
\eea
\bea
\lb{covmemory}
{\rm where}\  && \mathcal{M}(\tau,\tau^{\prime})
:= \frac{1}{2} \text{Cov}_{\CF }
\left(\frac{\hth(\tau,x^A)}{E(\tau,x^A)} \, , \,
\frac{\hth(\tau^{\prime},x^A)}{E(\tau^{\prime},x^A)}\right)  
\eea
defines the \textit{memory function} for the observer's past
null cone -- \textit{cf.} the definition (\ref{productav}) --
memorizing the spatial inhomogeneities and auto-correlation
properties in the expansion rate and in
the energy function of the incoming geodesic null rays encountered
along the path from the source to the observer. We have used that
\bea
\lb{dAint}
\ln(d_A)  = \frac{1}{2} \int^{\lambda}
\hth(\lambda^{\prime},x^A)\,{\mathrm d}\lambda^{\prime}
= \frac{1}{2} \int^{\tau}
\frac{\hth(\tau^{\prime},x^A)}{E(\tau^{\prime},x^A)}\,
{\mathrm d}\tau^{\prime} \ ,
\eea
where integration is along each geodesic null ray labelled
by comoving coordinates $x^A$, together with
Eqs.~(\ref{areadistevol}) and (\ref{tauparam}). 
The memory of shear and Weyl curvature along the
observer's past null cone is encoded in the shape of the
boundary $\partial\CF$ of the compact light front domain $\CF$,
according to Eqs.~(\ref{boundaryevolution}) and (\ref{exppath}). 

The time-range of dependence is expected to be determined by the
size of typical matter structures in the cosmological spacetime. 
Differentiating Eq.~(\ref{avareadistevol}) and using
Leibniz' rule leads to the area-averaged logarithmic focusing
equation:
\begin{subequations}
\bea
\lb{secondderiv}
\ptl^2_{\tau}\la \ln(d_A) \ra_{\CF}
&=& \ptl_\tau \mathcal{H}_\CF + \mathcal{M}(\tau,\tau)
+  \int^{\tau}\ptl_{\tau} \mathcal{M}(\tau,\tau^{\prime})\,
{\mathrm d}\tau^{\prime}  \ ,
\eea
with
\bea
\ptl_\tau \mathcal{H}_\CF
= -  \mathcal{H}^2_{\CF}  -  4 \pi \La \varrho  \Ra_{\CF} 
+ \mathcal{Q}^{\mathbf k}_{\CF} \;
+ \;{\mathcal S}^{\mathbf k}_\CF \ . 
\eea
\end{subequations}
The area-averaged logarithmic version of the reciprocity
relation (\ref{luminosityd2}) reads:
\be
\lb{avreciprocity}
\la \ln(d_L) \ra_{\CF} = 2 \la \ln(1+z) \ra_{\CF}
+  \la \ln(d_A) \ra_{\CF} \ ,
\ee
whose evolution along the observer's past null cone can be
determined through relations (\ref{aveq4}) and
(\ref{avareadistevol}). There is a sense in which the
logarithmic variables $\ln(1+z)$, $\ln(d_A)$ and $\ln(d_L)$
are natural for discussing area-averaged null signal propagation,
as the logarithmic transformation simplifies evolution
equations and preserves the local form of the reciprocity theorem.
We note that the distance modulus is defined as a linear
function of $\ln(d_L)$ by $\mu := (5 / \ln(10))
\ln(d_L / 10~\text{parsecs})$.

\medskip
We now consider area-averaging of the cosmological redshift
drift relation (\ref{zdrift}) over sources sprinkled continuously
and uniformly in volume over the $2$-dimensional screen space:
\be
\lb{zdriftav}
\hspace*{-0.0cm} \La \frac{{\mathrm d} z }{{\mathrm d}\tau_0}
\Ra_{\CF}
= \la 1+z \ra_{\CF}\,\alpha\,\bigg\rvert_{P}
- \la 1+z \ra_{\CF}\,\frac{E^\prime}{E}\,\bigg\rvert_{P}
- \La \alpha \Ra_{\CF}
+ \La \frac{E^\prime}{E} \Ra_{\CF} \ . 
\ee
Spatially inhomogeneous and anisotropic contributions enter the
cosmological redshift drift, also when area-averaged over
many sources. 
In particular, systematic effects are expected to be introduced
through the terms evaluated at the point of observation $P$, while
non-cancelling effects from structure along the individual null
rays might also play a role; see
Refs.~\cite{Heinesen:2020pms,Heinesen:2021nrc} for detailed
investigations of systematic effects entering the cosmological
redshift drift signal. 

\subsection{FLRW limit} 
\label{sec:FLRW}
It is worth considering the FLRW limit of the area-averaged
equations (\ref{areaexpansion}), (\ref{areaacceleration}),
(\ref{avconstraint2}) and (\ref{integrability}).  
We consider the FLRW spacetime metric written in hyper-spherical
local coordinates $\{t,r,\theta,\phi\}$ with line element,
\be
\lb{FLRWmetric}
{\mathrm d} s^2 = -{\mathrm d}t^2 + a(t)^2({\mathrm d}r^2
+ S^2_{k}(r) {\mathrm d}\Omega^2) \ , \qquad
S_{k}(r) := \sqrt{-k}^{-1} \sinh(\sqrt{-k} r) \ , 
\ee 
where ${\mathrm d}\Omega^2 = {\mathrm d}\theta^2
+ \sin^2(\theta) {\mathrm d}\phi^2$ is the solid angular
element on the unit sphere, and $a(t)^2 S^2_{k}(r)$ is the adapted
area measure on a sphere of proper radius $a(t)r$ contained in a
spacelike $3$-surface labelled by $t$. 
The dimensionless scale factor $a(t)$ is the conformal scaling of the static metric, 
and thus describes the temporal dependence of the spatial volume in the comoving frame; 
we employ the usual convention that $a(t_0) = 1$, where $t_0$ is the proper time
elapsed since the big bang singularity at the present epoch
(``here and now''). 
The spatial coordinates $\{r,\theta,\phi\}$ are comoving with a central
observer, where $\{\theta,\phi\}$ are angular coordinates
describing directions on the observer's sky, and $r$ is a
radial coordinate with dimension of length from which we can
get the proper geodesic distance $a(t)r$ 
away from the observer at an instant of proper time $t$. 
The curvature of the 3-dimensional spatial sections $k$ has
dimensions of inverse length scale squared, and the spatially flat
FLRW model is obtained when $k \rightarrow 0$. 
The observer's time variable $t$ is synchronous
with the proper time function $\tau$ associated with the 
irrotational $4$-velocity $u^\mu$ of the matter congruence,
of \textit{uniform} rest mass density $\varrho(t)$. 
The incoming geodesic null rays which constitute the central
observer's past null cone are per construction of the local
coordinates propagating along paths with
$\{\theta = \text{constant}, \phi = \text{constant} \}$. We
thus have in this reference frame:
\be
\lb{FLRWuke}
u^{\mu} = \delta_{t}{}^{\mu} \ , \qquad
e^{\mu} = \frac{1}{a(t)}\,\delta_{r}{}^{\mu} \ , \qquad 
k^{\mu} = E \left(\delta_{t}{}^{\mu} - \frac{1}{a(t)}\,
\delta_{r}{}^{\mu} \right) \ .
\ee 
The kinematic variables associated with $u^\mu$ and $e^\mu$ and
the energy function~$E$ of the geodesic null congruence
-- obtained from the connection associated with the line element
(\ref{FLRWmetric}) and the differential equation
(\ref{opticalenergy}) for $E$ -- are:
\be
\lb{FLRWdyn}
\theta = 3\,\frac{\dot{a}(t)}{a(t)} \ , \quad
\sigma_{\mu \nu} = 0 \ , \qquad 
\overline{\theta} =  \frac{2}{a(t)}
\frac{\partial_{r} S_k(r)}{S_k(r)} \ , \quad
\overline{\sigma}_{\mu \nu} = 0 \ , \qquad
E=\frac{ \mathcal{E}(L) }{a(t)} \ ,
\ee
where $\dot{}:= \partial_t$, and
$L := \int^t {\mathrm d}t^{\prime}/a(t^{\prime}) + r$
is constant over each past null cone of the central observer. 
The function $\mathcal{E}(L)$ is an initial condition for the
geodesic null congruence and might be chosen in accordance with
the observational frequency of interest; \textit{cf.} section
\ref{sec:nullcone}. 
The null scalar variable of the geodesic null congruence is given
by
\be
V(L) =  - \int^L  \mathcal{E}(L^{\prime})\,
{\mathrm d}L^{\prime} \ ,
\ee
which can be verified by computing its
gradient and recovering $k^\mu$.

The kinematic variables associated with $k^\mu$ and $l^\mu$
follow from plugging the expressions (\ref{FLRWdyn}) into
the relations (\ref{kinhat}) and (\ref{kintilde}), yielding:
\begin{subequations}
\bea
\lb{FLRWdynk}
\frac{\hth}{E} & = & \frac{2}{3}\,\theta 
- \overline{\theta} = 2\,\frac{\dot{a}(t)}{a(t)}
- \frac{2}{a(t)}\, \frac{\partial_{r} S_k(r)}{S_k(r)} \ , \qquad 
\hsig_{\mu \nu} = 0 \ , \\ 
\frac{\tith}{E} & = & \frac{2}{3}\,\theta  + \overline{\theta}
=  2\,\frac{\dot{a}(t)}{a(t)}
+ \frac{2}{a(t)}\,\frac{\partial_{r} S_k(r)}{S_k(r)} \ , \qquad 
\tisig_{\mu \nu} = 0 \ . 
\eea
\end{subequations}
The contribution $\pm 2\partial_{r}(S_k(r))/S_k(r)/a(t)$ to the
area expansion rates accounts for the change of the area measure as
the proper radius of the $2$-dimensional screen space is decreased
or increased, respectively, when propagating the screen towards or
away from the vertex point $P$ of the observer's past null cone
along the incoming/outgoing null rays. This contribution
becomes singular at the vertex point $P$ at coordinate value
$r = 0$, where the angular measure of the $2$-dimensional screen
space tends to zero.

We now consider the area-averaged equations
(\ref{areaacceleration}, \ref{avconstraint2}), in the limit of
the FLRW cosmologies. In this limit, the backreaction terms 
$\mathcal{Q}^{\mathbf{k}}_{\CF}$ and $\mathcal{Q}^{\mathbf{k}
\mathbf{l}}_{\CF}$ and the right-hand
side of Eq.~(\ref{rhoav}) vanish. The Ricci scalar of the
embedded $2$-surfaces is
\bea
\lb{curvFLRW}
{}^{(2)}\!R =  \frac{{}^{(3)}\!R }{3}
+ \frac{1}{2}\, \overline{\theta}^2 \ ; \qquad 
{}^{(3)}\!R = \frac{6 k}{a^2(t)} \ , 
\eea
where ${}^{(3)}\!R$ is the 3-Ricci curvature scalar associated with
the canonical FLRW spacelike $3$-surfaces ${\cal S}(t)$.
Using these results, along with Eqs.~(\ref{areaexpansion}) and
(\ref{FLRWdynk}), on both sides of the area-averaged Gauss
embedding constraint~(\ref{avconstraint2}) gives:
\bea
\lb{avconstraint2FLRW}
\left(\frac{\dot{a}}{a} \right)^2
+ \frac{1}{4}\,\overline{\theta}^2
- \frac{\dot{a}}{a}\,\overline{\theta}
=  \frac{8 \pi}{3} \varrho  + \frac{1}{3}\Lambda
- \frac{ k }{a^2} -  \frac{\dot{a}}{a}\,\overline{\theta}
+ \frac{1}{4}\,\overline{\theta}^2  \ ,\nonumber
\eea
where it has been used that all area-averaged variables are
constant over the $2$-dimensional screen space. The terms
involving $\overline{\theta}$ cancel, and we arrive at
the first of Friedmann's equations:
\bea
\lb{avconstraint2FLRW2}
\left(\frac{\dot{a}}{a} \right)^2
=  \frac{8 \pi}{3} \varrho  + \frac{1}{3}\Lambda
- \frac{ k }{a^2} \ .
\eea
The Raychaudhuri equation can be recovered by noticing that
\begin{subequations}
\bea
\lb{FLRWacc}
E^{-1}k^{\rho}\partial_{\rho}\left( \frac{\hth}{E} \right)
& = & (u^{\rho} - e^{\rho})\partial_{\rho}
\left(2\,\frac{\dot{a}}{a} -  \overline{\theta} \right)
= 2\,\frac{\ddot{a}}{a} - 2 \left(\frac{\dot{a}}{a} \right)^2
- \dot{\overline{\theta}} + \frac{1}{a}\,\overline{\theta}^\prime
\ , 
\eea
\bea
\lb{phideriv}
{\rm where}\qquad\  \dot{\overline{\theta}}
= -  \frac{\dot{a}}{a}\, \overline{\theta} \ ,  \qquad
\overline{\theta}^\prime = -\frac{2k}{a}
- \frac{a}{2}\,\overline{\theta}^2 \ , 
\eea
\end{subequations}
follows from the definition of
$\overline{\theta}$,~Eq.~(\ref{FLRWdyn}). This gives for the
acceleration equation (\ref{areaacceleration}):
\bea
\lb{sfhthaccFLRW}
\frac{\ptl^2_{\tau}a_{\CF}(\tau)}{a_{\CF}(\tau)}
&=&  \frac{\ddot{a}}{a}
- \frac{1}{2}\,\frac{\dot{a}}{a}\,\overline{\theta}
- \frac{k}{a^2} \ .  
\eea
We also have from Eq.~(\ref{areaacceleration}) that
\bea
\lb{sfhthacc2FLRW}
 \frac{\ptl^2_{\tau}a_{\CF}(\tau)}{a_{\CF}(\tau)}
= - 4 \pi  \varrho    +  \left(\frac{\dot{a}}{a} \right)^2
- \frac{1}{2}\,\overline{\theta}\,\frac{\dot{a}}{a}
=  -  \frac{4 \pi}{3}  \varrho  + \frac{1}{3}\Lambda
- \frac{ k }{a^2}
- \frac{1}{2}\,\overline{\theta}\, \frac{\dot{a}}{a} \ , 
\eea 
where it has been used that
\bea
\lb{FLRWEderiv}
 \frac{\hth}{E}\,\frac{E^{-1}k^{\rho}\partial_{\rho}E}{E}
= - 2\left(\frac{\dot{a}}{a} \right)^2
+ \overline{\theta}\, \frac{\dot{a}}{a} \ .
\eea
Finally, equating Eqs.~(\ref{sfhthaccFLRW}) and
(\ref{sfhthacc2FLRW}), we arrive at Friedmann's acceleration law:
\bea
\lb{sfhthaccFLRWfinal}
\frac{\ddot{a}}{a} =   - \frac{4 \pi}{3} \varrho
+ \frac{1}{3}\Lambda \ . 
\eea
It is illustrative to derive the scaling of the
Gaussian curvature scalar of the FLRW screen spaces: 
\bea
\lb{GcurvFLRW}
K = \frac{1}{2} {}^{(2)}\!R  = \frac{ 1 }{a^2(t) S^2_k(r) }
\left[\, k S^2_k(r) +  \left( \partial_{r} S_k(r) \right)^2
\,\right]
= \frac{ 1 }{a^2(t) S^2_k(r) } \ .
\eea
This expression is manifestly positive when both 
$a(t)$ and $S_{k}(r)$ are non-zero. The
first equality in Eq.~(\ref{GcurvFLRW}) follows from
Eqs.~(\ref{curvFLRW}) and (\ref{FLRWdynk}), while the second
follows from the definition of $S_k(r)$ and hyperbolic identities.
The FLRW expression for the Gaussian curvature scalar
(\ref{GcurvFLRW}) is nothing but the inverse area measure of the
light fronts. The integral of Eq.~(\ref{GcurvFLRW}) over the full
light front yields the area of the unit sphere, $4\pi$, consistent
with the Gauss--Bonnet theorem for closed surfaces
(\ref{GBclosed}). For light front subdomains,  the integral of $K$
is simply the constant area covered by the subdomain of the unit
sphere. 

\section{Discussion}
\label{sec:discussion}
This paper offers an area-averaging formalism for describing
the evolution of dynamical variables and 
observable quantities along the observer's past null cone in the
setting of generic irrotational dust cosmologies with a
cosmological constant. This formalism applies to all kinds of null signals, e.g., those
carried by photons, but also a future gravitational wave sky could
be examined using this formalism. The formalism is based on
the assumption that the cosmological dynamics progresses
sufficiently smoothly so that no caustics will form on the past
null cone of the observer.  

The area-averaged macroscopic system of equations on 
$2$-dimensional light fronts derived in
section~\ref{sec:avevolution} involves two independent
backreaction terms. The system of light front averaged equations
comprises a larger set of global variables than the analogous
volume-averaged macroscopic system of equations adapted to
the spacelike $3$-surfaces given in Ref.~\cite{buchert:grgdust}.
This is to be expected, as
additional dynamical variables are introduced when the dynamics of
the null cone generating congruence is included.
Since the area-averaging operation is adapted to the observer's
past null cone, there is a flow of matter world lines across the
averaging domain, which further complicates the light front
averaged equations.  
We note that caution is required when invoking simplifying
assumptions. A radical example would be to get rid of all
of the shear variables of the problem by simply setting the
area shear rate of the geodesic null congruence and the
volume shear rate of the matter congruence to zero. 
However, such approximations at the local level of the
cosmological spacetime turn out to be extremely restrictive;
see Ref.~\cite{Ellis:2011} and references therein. If, for
instance, the volume shear rate of the matter congruence
is required to be zero, $\sigma_{\mu \nu}=0$, then
-- since its vorticity is already required to be zero
-- this implies that the dust cosmology considered must be 
of the FLRW class~\cite{ell1967}. 
Vanishing of the area shear rate of the geodesic null congruence,
on the other hand, implies that the evolution of this null
congruence is affected only by the matter
(hence, Ricci curvature) it encounters,\footnote{The
Weyl curvature has no influence on the properties of the
geodesic null congruence in this case.}
and that this null congruence must be a principal
null direction of the Weyl curvature tensor; this fact might be
viewed as a generalization of the Goldberg--Sachs theorem for
vacuum spacetimes \cite{Ellis:2011,GoldbergSachs}.

The generality of the presented formalism leads, by construction, to a set of balance equations that do not form a closed set. Quantification of the level of backreaction is intimately related to the implementation of closure conditions for the screen-space averaged variables. Concrete inhomogeneous models at the level of \textit{area-averaged variables} can be studied in order to arrive at appropriate statistical descriptions for the propagation
of geodesic null ray bundles in spatially inhomogeneous dust cosmologies. 
One might for instance employ exact solutions such as 
Lema\^\i tre--Tolman--Bondi solutions or the more general classes of Szekeres solutions.
They can be employed to design `Swiss cheese models', 
within which light propagation has already been studied
in great detail; see \cite{Lavinto:2013exa,Koksbang:2021zyi} for
recent investigations into observational backreaction effects in
Swiss cheese models.
Generic structure formation models, based on perturbative assumptions and not restricted by symmetries, are a natural next step of investigation.
In the present setup of the irrotational matter congruence, appropriate models have been constructed that are based on 
relativistic perturbation theory combined with exact averages, e.g. \cite{lischwarz}, and 
the relativistic generalization of Lagrangian perturbation theory, recently reviewed in \cite{Universe}.\footnote{See the references therein, also to earlier work on other relativistic perturbation theories, the history of Lagrangian perturbation theory, and e.g. \cite{RZA_2} for their application as a closure condition for the $(3+1)-$averaged system.} 
Ongoing work investigates a non-perturbative generalization of these schemes, controlled by the Szekeres class of solutions and that contain the most general Szekeres solutions as the exact body \cite{GRZA}. 

Studies of exact scaling solutions of the area-averaged system could also provide
more insight, similar to those employed in the 
$3+1$ slicing formalism \cite{roy:instability}, and they can 
be explored to provide closure conditions for the area-averaged equations.  
The scaling behaviours suggested by the integrability
condition \eqref{integrabilitycompact} at first glance are those
where the relevant variables are uncoupled, but we expect from
Ref.~\cite{roy:instability} that generic scaling solutions
dynamically couple the area-averaged variables. 
In this context, topological closure conditions might be relevant: an insight from the Gauss--Bonnet
theorem applied to light fronts shows that there is no such
coupling of the scalar curvature of light fronts to other dynamical
variables for all-sky
averages, unlike in the $3$-dimensional case:
the area-averaged scalar parts of Einstein's field equations
on the light fronts simplify through the Gauss--Bonnet theorem
when the total (all-sky) light front is considered, or when the
curvature of the embedding of the boundary is associated with a
conserved current, as detailed in section \ref{sec:GB}. 
Integral-geometric properties analogous to the Minkowski
functionals in Euclidian spaces appear naturally when all-sky averaging
over light fronts is performed. Hence, it is for the case of \textit{all-sky averages},
where  we could expect simplifications after evaluating global
contributions on topologically closed light fronts. We 
do not in general expect the light front backreaction terms to
vanish globally.  However, in cases where those terms are covariant
divergences, a closed space would  erase them, as in the case 
of flat light fronts (corresponding  insights have been
developed in Newtonian cosmology \cite{buchertehlers}).

Another arena of application of the formalism is that of general relativistic numerical simulations. 
Cosmological simulations have already been investigated in the context of backreaction and observables;
cf. \cite{Giblin:2015vwq,Giblin:2016mjp,Bentivegna:2015flc,Adamek:2015eda,East:2017qmk,East:2019chx,Macpherson:2018btl} for recent works. In these works, cosmological backreaction was generally found
to be small at large scales where observables were also found to
agree well with the FLRW prediction, whereas on smaller scales
Newtonian simulations of structures combined with relativistic
ray tracing could account for distortions of the FLRW background
observations. The FLRW metric has thus proven itself robust
towards the initial perturbations employed within these simulated
universe scenarios. This robustness might seem surprising given
the highly non-linear structures which are allowed to develop in
these codes. However, this may be related to the fact that a foliation that
inherits the properties of the longitudinal gauge of standard
perturbation theory is present throughout the cosmic evolution of
these simulations despite the presence of non-linear
structures \cite{Giblin:2018ndw,Clifton:2020oqx}. Furthermore, 
the global architecture remains that of Newtonian cosmology of a flat 
$3$-torus topology, which forces the global model to evolve according to
the assumed background cosmology. 
These are conjectures that should be examined in future works,
together with the question of which type of initial perturbations
and topological constraints may cause the FLRW solution to
be globally unstable as a model for the average evolution, 
an instability that generically occurs in a dynamical system analysis of 
the averaged $3+1$ system \cite{roy:instability}, pointing to negative average curvature
on large scales \cite{buchertcarfora:curvature}. The formalism developed in this paper sheds light on
the conditions that must be satisfied for backreaction to be
present in the context of observables, and may lead to interesting
developments on simulated spacetime scenarios with backreaction. 

Turning to observables, the evolution equations for the area-averaged effective
angular diameter distance derived in section \ref{sec:averageobs}
are partly expressed in terms of the area-averaged light front
variables introduced in section \ref{sec:avevolution}, but involve
additional terms expressed through the memory function
$\mathcal{M}(\tau,\tau^{\prime})$, for which one needs to prescribe
a model to close the area-averaged macroscopic system of
equations. Backreaction in the area-averaged effective
angular diameter distance might arise due to
\sayy{direct backreaction} through the memory function or 
$\mathcal{Q}^{\mathbf k}_{\CF}$, or through \sayy{indirect
backreaction}
causing both $\mathcal{H}_{\CF}$ and $ \La \varrho
\Ra_{\CF}$ to evolve differently than in the Friedmannian case.
The evolutions along the observer's past null cone of the
area-averaged cosmological redshift and of the area-averaged
cosmological redshift drift both induce additional terms which
must be dealt with -- for detailed investigations of the
expression for cosmological redshift drift in a generic
cosmological spacetime, and the effect of regional spatial
inhomogeneities and anisotropies on the cosmological redshift
drift signal, see Refs.~\cite{Heinesen:2020pms,Heinesen:2021nrc,Heinesen:2021qnl,Korzynski:2017nas}. 

We emphasize that care must be taken when using the
area-averaged relations to interpret cosmological data sets. 
Some challenges for interpreting the theoretically given
area-averaged observable quantities include
(i)~identifying spacelike $3$-surfaces of constant proper time
on which the area-averaging operation is formulated;
(ii)~relating the effective area-averaging operation
employed when observing many emitters over the sky to the
volume-averaging operation. Emitters might not sample the
area of the emission screen space fairly, and null signals will
interact with the matter of the spacetime;  
(iii)~incorporating the properties of the observer's local
vicinity relative to that of an observer with a volume-unbiased
view of the Universe. Here follows a discussion of these
challenges. 
\begin{itemize}
\item[(i)] Relative age measurements \cite{JimenezLoeb} might
serve to distinguish spacelike $3$-surfaces of constant proper
time, although the errors associated with current relative age
measurements are of the order of 20\% \cite{Ratsimbazafy}. 
The foliation of the observer's $3$-dimensional past null cone is
not unique, and the choice to foliate it in terms of a family of
spacelike $3$-surfaces ${\cal S}:\!\{\tau=\text{constant}\}$
of constant proper time with respect to the irrotational 
matter congruence could be modified, which would change the
dynamical system of equations accordingly. 
It has been proposed to consider level
surfaces of constant cosmological redshift $z$ as an
observationally relevant foliation of the observer's past
null cone \cite{Fanizza2019pfp}.
Regions of constant cosmological redshift are identifiable to a
high precision, and are therefore observationally preferred.
However, when $z$ is not a monotonic function along null paths,
this destroys the otherwise
observationally intuitive notion of $z$ as a foliation scalar.
In situations where the effective fluid
matter model is describing scales well above that of collapsing
structures, then the redshift function might be re-established as
a parametrization of light fronts. It must be assessed
in the physical situation at hand whether the cosmological
redshift function is indeed a hypersurface-forming scalar. 
In the general case, where the cosmological redshift function is
not hypersurface-forming, we might, nevertheless, average over
regions of constant cosmological redshift, without these regions
being well-defined as causally ordered and non-intersecting
submanifolds. 

\item[(ii)] Luminous astrophysical sources emitting
electromagnetic radiation or gravitational wave signals are
in general not uniformly distributed in volume over the Universe.
Such sources are concentrated in over-densities of the matter
distribution, and in this regard an observer should be observing
a mass-biased picture of the Universe. On the other hand, after
being emitted from a source, the null signals, which a given
observer receives, tend to propagate in empty space (since signals
propagating in regions with matter are likely to be scattered
or absorbed by other particles, and, therefore, simply do
not arrive at the observer's position). Such subtleties must
ideally be taken into account in an observationally matched
averaging procedure. However, correctly accounting for such
biases would be involved and the
correction procedures would depend on the type of cosmological
probe. As a first approximation we might simply assume an ideal
situation where sources and null signals are probing the
volume-averaged cosmological spacetime. 

\item[(iii)] All cosmological measurements are dependent on the
properties that hold in the immediate vicinity of the small
segment of a single world line from which we observe the
Universe. Whereas cosmological observations are directly or
indirectly averaged over the emitter positions when performing
statistical analysis with many cosmological data points, the same
does not apply to the observer's position. Some
cosmological measurements might be sensitive to the
observer's position, whereas others might be subject to less bias. 
\end{itemize}
The area-averaging formalism provided in this paper may be combined
with appropriate statistical assumptions in order to formulate an
average dynamical theory for dynamics and observables on our past
null cone.

Synthetic multi-variate cosmological data on the joint
distribution of luminous astrophysical sources and spatial
geometry inside and on a modelled observer's past null cone
may be generated by simulation and analyzed by means of methods
from inductive statistical inference;
\textit{cf.} the textbook introductions,
Refs.~\cite{mce2020,geletal2014}, and \cite{hve2020}. This
approach allows for the possibility of factoring in all information
that is relevant to dealing systematically with the uncertainty
pertaining to a quantitative problem at hand. In this way (and
when linked to exact cosmological dust solutions to begin with),
a set of interval estimates for area-averaged observables such as,
e.g., the cosmological distance measures may be calculated for
specific configurations and then compared to standard FLRW-based
values for these quantities. The amount of available observational
data to be integrated in this process is steadily increasing. On
the technical side, a long term project of this kind of
cosmological data analysis may be informed by and validated through
supervised solution algorithms developed in contemporary machine
learning such as neural networks; \textit{cf.}, e.g., the textbook
introduction Ref.~\cite{ng2018}, and examples on applications in
Refs.~\cite{gabetal2018,machinelearning:lightcone}.  

Our analytical results can be profitably employed for numerical simulations
of null cones. As mentioned in the discussion above, it is
important to examine the regime of global stability of the FLRW solution
for predicting observables by means of numerical relativity. 
Relativistic effects in light propagation have been computed in
linearized weak-field approximations of general relativity, while
using (N-body or hydrodynamic) Newtonian simulations as input to
generate the matter distribution of the universe model
\cite{Borzyszkowski:2017ayl,Breton:2018wzk}. Light rays have also
been tracked fully relativistically through idealized but
non-trivial inhomogeneous and anisotropic model universes with backreaction 
\cite{Lavinto:2013exa,Koksbang:2019glb,Koksbang:2020zej} and
within post-Newtonian model universes
\cite{Sanghai:2017yyn,Grasso:2021zra}. 
Weak-field relativistic N-body simulations, which allow to treat
certain perturbative modes non-linearly, have been employed in
order to model both the matter distribution and the null cone of
an observer \cite{Adamek:2018rru}. 
Null cones have been generated in general relativistic simulations
codes; however, only tracking the evolution in the linear
regime of density contrasts \cite{Giblin:2016mjp}. See also
\cite{East:2017qmk} for a study comparing observables in Newtonian
N-body codes and general relativistic hydrodynamic simulations of
structure formation, while employing relativistic ray-tracing in
both cases. Finally, ray tracing codes for the 
general relativistic and non-linear universe simulations in
Refs.~\cite{Macpherson:2016ict,Macpherson:2018btl} have
recently been developed \cite{Macpherson:2022eve} with exciting
applications to relativistic modelling of cosmic observables
\cite{Macpherson:2022}. 

The framework developed in this paper is complementary to
model-independent cosmographic frameworks first considered in
detail by \cite{krisac1966,1985PhR...124..315E} and more recently
by \cite{Clarkson:2011uk,Heinesen:2020bej,Heinesen:2021nrc,Heinesen:2021qnl},
with applications to numerical simulations in
\cite{Macpherson:2021gbh,Heinesen:2021azp}. While the cosmography
developed in these works is powerful for extracting cosmological
information from data without the assumption of a cosmological
metric or a dynamical theory, and primarily apply to  measurements
made at low redshifts, the framework considered in this paper
considers the dynamical evolution and theoretical average of light
fronts as they propagate towards the observer from astrophysical
sources that may be distant. The cosmographic frameworks in
\cite{Heinesen:2020bej,Heinesen:2021qnl} are furthermore concerned
with the angular dependence of observables over the sky of the
observer, while the framework developed here is concerned with the
average over (a patch of) the sky of the observer,  which is applicable in general for finite-area screen
spaces with or without boundary. Future investigations may
consider a link to the analyses of galaxy catalog data on shells around the observer or Cosmic
Microwave Background radiation data that are often investigated
globally on topological all-sky support manifolds
with the help of integral-geometric measures like the Minkowski functionals.  
\begin{acknowledgements}
This work is part of a project that has received funding from the
European Research Council (ERC) under the European Union's
Horizon 2020 research and innovation programme (grant agreement
ERC advanced grant 740021--ARTHUS, PI: TB).
The authors would like to thank J\"{u}rgen Ehlers, George Ellis, Pierre
Mourier, Dominik Schwarz and Nezihe Uzun for useful discussions, and 
Giuseppe Fanizza, Syksy R\"as\"anen and Dennis Stock for valuable comments on the manuscript.  
We would in addition like to thank the anonymous referee for constructive suggestions that helped improving the paper.  
This work has been begun during a visit of TB in 2007 to the
University of Bielefeld, Germany. TB wishes to thank Dominik
Schwarz for his invitation to hold a temporary C4-chair at the
department of physics. HvE acknowledges the generous hospitality of
the UCT Cosmology and Gravity Group during the period from July to
October 2010 when a significant share of the work underlying this
paper was accomplished.
\end{acknowledgements}
\appendix
\renewcommand{\theequation}{\thesection.\arabic{equation}}
\setcounter{equation}{0}
\section{Light front-adapted spacetime metric}
\label{appA}
We shall consider the embedding of the light fronts into the
cosmological spacetime, and write the metric tensor in local
coordinates that are adapted to the screen space, 
$x^\mu = (\tau, V, x^A)$. We require that the local coordinates
$x^{A}$, with $A = 2,3$, satisfy the propagation law
$k^{\rho}\partial_{\rho}x^A = 0$. The propagation laws for
$\tau$ and $V$ are also fixed through
$k^{\rho}\partial_{\rho}\tau = E$ and
$k^{\rho}\partial_{\rho}V = 0$ (see Eqs.~\eqref{tauparam} and
\eqref{eq:kproperties2}), and the full adapted system of local
coordinates $x^\mu = (\tau,V, x^A)$ is thus specified throughout
the past null cone domain, once initial conditions for the local
coordinates are fixed at a screen space. In these local
coordinates, we have that
\begin{align}
\lb{kucomponents}
&k_\mu = (0,  {1},  0,0) \ ;  && u_\mu = ( {-1},0,0,0)   \ ; \\ 
&k^\mu = ( {E},0,  0,0) \quad  ; &&u^\mu = ( {1, - E, EU^A}) \ , 
\lb{kucomponentsraised}
\end{align}
where $U^A := u^{\rho}\partial_{\rho}(x^A) /E$ defines the drift
of the screen space coordinates in the matter frame. The one-form
components in \eqref{kucomponents} follow directly from the
definitions $k_\mu := \partial_\mu V$ and
$u_\mu :=  - \partial_\mu \tau$. The vector components in
\eqref{kucomponentsraised} follow from the definition of the
energy function $E := - k^\mu u_\mu$ and the transport rules
(shift vectors) $k^{\rho}\partial_{\rho}(x^A) = 0$ and
$u^{\rho}\partial_{\rho}(x^A) = E U^A$. We may now write the
projection tensor onto the light fronts \eqref{projection} in the
adapted local coordinate system $x^\mu = (\tau, V, x^A)$:
\be
\label{pGLCmetric}
 p_{\mu\nu} =
\left(
\begin{array}{ccc}
0 & 0 & \vec{0} \\
0 & \, \, U^2 \,  \, & U_B \\
\vec{0} & U_A & p_{AB}  \\
\end{array}
\right) \ ;~~ ~~~~~
p^{\mu\nu} =
\left(
\begin{array}{ccc}
0 & 0 & \vec{0}   \\
0 & \, \, 0 \, \, & \vec{0}  \\
\vec{0} & \vec{0} & p^{AB} \\
\end{array}
\right) \ , 
\ee
where $U_A := p_{AB} U^B$ and $U^2 := p_{AB} U^A U^B$, and where
the area-adapted screen space metric $p_{AB}$ has inverse $p^{AB}$.  
The tensor components in \eqref{pGLCmetric} follow from the
orthogonality conditions $p_{\mu \nu} k^\nu = p_{\mu \nu} u^\nu
= 0$ and $p^{\mu \nu} k_\nu = p^{\mu \nu} u_\nu = 0$, respectively. 
Note that in general the values of the components $p_{11}$,
$p_{1A}$ and $p_{A1}$ are non-zero, which comes from generally
non-zero values for $u^1$ and $u^A$ in \eqref{kucomponentsraised}. 

Using the definitions \eqref{def:kdef} and \eqref{projection}, we
might formulate the metric tensor for the cosmological
spacetime as $g_{\mu \nu} = k_\mu k_\nu/E^2
- (k_\mu u_\nu + u_\mu k_\nu)/E + p_{\mu \nu}$,
and insert Eqs.~\eqref{kucomponents}, \eqref{kucomponentsraised},
and \eqref{pGLCmetric} in this formulation to obtain 
\be
\label{GLCmetricappendix}
{ g_{\mu\nu} =
\left(
\begin{array}{ccc}
0 & 1/E & \vec{0} \\
1/E & \, (1/E^2) \!  + \! U^2 \, & U_B \\
\vec{0} &  {U_A} & p_{AB}  \\
\end{array}
\right) \ ;~~ ~~~~~
g^{\mu\nu} =
\left(
\begin{array}{ccc}
-1 & \, E \, &  {-} E U^B \\
E & 0 & \vec{0}  \\
 {-} E U^{A} & \vec{0} & p^{AB} \\
\end{array}
\right) \ .}
\ee

\medskip
\noindent
\textbf{Data Availability statement:}

\noindent
Data sharing not applicable to this article as no datasets were generated or analysed during the current study.
%

\end{document}